# Supercooling-enabled giant and tunable thermal rectification ratio of a phase change thermal diode

*Zhaonan Meng, Raza Gulfam, Peng Zhang\*, Fei Ma*
Institute of Refrigeration and Cryogenics, Shanghai Jiao Tong University, No. 800 Dongchuan Road, Shanghai 200240, China
E-mail: zhangp@sjtu.edu.cn

Phase change thermal diodes (PCTD) suffer from fairly low thermal rectification ratio, which hampers their widespread utilization as thermal management and control units for cutting-edge technologies, encompassing photovoltaics, thermoelectric modules, batteries and other miniaturized electronic products, etc. Herein, a PCTD prototype is theoretically designed via heat conduction and convection models, fabricated and experimentally executed, accessing an unprecedentedly giant thermal rectification ratio of 3.0 at ambient temperature. Theoretical model additionally introduces freezing temperature to address supercooling-enabled extended-convection and liquid-height-dependent effective thermal conductivity. With optimized length ratio and feasible assembly of two phase change terminals, the emerging physical states of thermal media continuously manipulate heat flux in the forward direction and reverse direction operations within temperature bias of 10~40 °C. The most significant finding discloses the fact that supercooling elongates the overall workable temperature bias range, while manual supercooling release allows to tune the thermal rectification ratio at any temperature bias within 10~33 °C. Integrating the conventional asymmetric thermal transport mechanism with state-specific heat transfer hysteresis helps establish the novel governing mechanism.

## 1. Introduction

The necessity of waste heat scavenging and mitigation of thermal emissions through environment-friendly as well as techno-economic scenarios are declared to be the universal goals. Therefore, sustainable heat recovery, either from uninterruptable waste heat sources, miniaturized electronic products or intermittent solar sources, via directional heat flux manipulation is emerging as one of the fundamental interests. In the paradigm of multi-oriented demands, thermal technology has been recently utilized to create innovative thermal diodes, which is expected to be employed for efficient thermal computing and information processing,[1-2] cryogenic radiative insulation,[3] satellite temperature control,[4-5] thermal logic circuitry,[6] and so on. By definition, thermal diode is a bi-terminal nonlinear thermal element which allows the heat flow in one direction at a forward temperature bias, while blocking the reversal heat flow at a reverse temperature bias. This unidirectional and preferential behavior is normally called thermal rectification. To further rectify the heat flow, a co-junctional thermal diode usually requires one material made up of asymmetrical structure[7-14] or two entirely different materials[15-19] with various temperature-dependent thermo-physical properties. Briefly, thermal diode is a unifunctional device which is capable of operating in a forward direction with heat flux $q_{for}$ and in a reverse direction with heat flux $q_{rev}$ at the same temperature bias, where $q_{for} \geq q_{rev}$. And the ratio of steady-state forward heat flux to reverse heat flux at a certain temperature bias is known as thermal rectification ratio $R$ which is always larger than or equal to 1, as given in Equation (1). Meanwhile, increasing the thermal performance, which is quantified and evaluated by the thermal rectification ratio, is still underway, and demands further novel working mechanisms and thermal media (working materials).

$$R = \frac{q_{for}}{q_{rev}} \tag{1}$$

The certain types of thermal media, i.e., phase variant or phase invariant, can generate diversified working mechanisms that are necessary for thermal rectification. For instances, phase variant (normally known as phase change materials and abbreviated as PCMs) thermal media (e.g., solid-solid and solid-liquid PCMs) allow heat conduction in solid state (lattice vibrations), and thermal convection in liquid state (buoyant motion), thus they can develop dual-nature thermal diodes based on double-phase-oriented heat transfer mechanism.[20,21] Phase invariant thermal media can undergo either conduction or radiation,[22,23] where single-phase-oriented heat transfer mechanism is created.

Dating back to 1936, Starr presented a solid rectifier and provided the insight of directional heat transfer at the junction of copper and cuprous oxide nanostructures, which led to disclose the innovative phenomenon of asymmetrical thermal conductance in relevance with electron theory of conduction.[24] Since then, a series of thermal diode materials have been investigated conforming to variable thermal conductance triggered by means of various heat carriers. However, indulgence of phonons has surprisingly rendered the governing mechanisms to be somewhat complicated in comparison with electron heat carries.[25] Most probably, in order to minimize the design complications due to high-frequency phonons, the utilization of two materials with almost similar phonon-bands is recommended. Such a system has been reported with thermal rectification ratio of 1.43, which is ascribed to temperature-dependent asymmetrical thermal conductivities of the interzones in the temperature range of 40-100 K.[26] Apart from achieving the asymmetrical thermal conductance in a single material via gradual change of external temperature, this mechanism can also be realized by introducing either structural or state-change phenomena. For instance, a thermal rectification ratio of 1.14, under a temperature bias of 2 K in forward and reverse bias operations, has been reported on behalf of structural phase change in $MnV_2O_4$ bar (top terminal) bonded with phase invariant $La_{1.98}Nd_{0.02}CuO_4$ bar (bottom terminal).[27] Moreover, two juxtaposition materials have been configured to work at quite high temperatures, for example, Al-based quasicrystals coupled with silver telluride ($Ag_2Te$) tend to operate between 300 K (cold end) and 543 K (hot end), resulting in a thermal rectification ratio of 1.63. Since a desirable thermal conductivity drop of $Ag_2Te$ occurs at around 420 K because of structural phase change, it laid down the basis that temperature lower than this limit would not yield a pronounced thermal rectification.[28] It is therefore concluded that the most probable reason of higher operational temperature biases of previously described bulk thermal rectifiers is ascribed to the nature of employed materials.

While in most of the applications, maintaining such a huge temperature bias may not be feasible, urging to explore low temperature bias-driven thermal media.

Although many kinds of thermal media have been employed to fabricate thermal diode terminals, the design complications, arose on the account of selected thermal media, have impeded thermal rectification ratio. For example, if thermal conductivity of the first terminal increases with the rising temperature and the second terminal also behaves in the same way, then the high thermal rectification ratio in either direction may not be achieved. Such a configuration composed of two phase change terminals has been analyzed, providing a thermal rectification ratio of only 1.23[29] and further suggesting to utilize alternative PCMs. In this context, PCMs are enormous and generally known on the basis of interconvertible physical states depending on the melting process or solidification process. However, before opting out a PCM, the fundamental categorization is imperative, which includes organic and inorganic materials. The best suited examples of organic PCMs are paraffin waxes[30] and non-paraffin materials such as fatty acids,[31] poly-alcohols[32] and certain polymers (e.g., PANIPAM (poly-n-isopropylacrylamide), polyurethane and polyethylene glycol, etc.).[33] While the inorganics include salt hydrates[34] and metallics.[35] Moreover, melting temperature ranges of PCMs have been classified by Zhang et al.,[36] i.e., low melting temperature range <100 °C (e.g., paraffin waxes), medium melting temperature range 100~300 °C (e.g., salt hydrates), and high melting temperature range >300 °C (e.g., metals/metallic alloys). The major benefit of PCMs is the variation of physical states with temperature, creating the rise or fall of thermal conductivity during every emerging state, which is an integral criterion to robustly modulate the heat flux either in the forward or reverse direction operation. Meanwhile, most of the PCMs also undergo a phenomenon like supercooling, which is however avoided for the PCMs being used to erect thermal energy storage and thermal management systems.[37] Shortly, need-based novel thermal diodes can be fabricated by taking full advantage of different melting temperatures followed by appropriate selection of PCMs. A novel phase-change-driven thermal diode has recently been fabricated using polystyrene foam impregnated with octadecane paraffin (PFH-O). The reported thermal rectification ratio of PFH-O thermal diode was up to 1.43, which was further enlarged to 2.6 upon coupling the PFH-O with PNIPAM aqueous solution.[20,38] In order to jump from the highest claimed thermal rectification ratio of 2.6 to the improved magnitudes, there is an urgent need to disclose a unique mechanism of PCM-based thermal diodes.

**Table 1.** Tentative summary of the experimental (E) and/or theoretical (T) results of the achieved maximum thermal rectification ratios ($R_{max}$) with respect to thermal diode materials and temperature of the cooler ($T_c$)/heater ($T_h$). The table is aligned according to the thermal rectification ratio $R_{max}$. The structural transition refers to a phase change within a single state (i.e., solid or liquid), while the state transition refers to a tranfomration of one state into another state (e.g., solid into liquid).

| Authors | Study type | Employed materials | Transition | $R_{max}$ | $T_c/T_h$ (°C) |
|---|---|---|---|---|---|
| W. Kobayashi[27] | E and T | $La_{1.98}Nd_{0.02}CuO_4/MnV_2O_4$ | Structural | 1.14 | -217.6/-215.6 |
| R. Chen[29] | E and T | Eicosane/PEG4000 | State | 1.23 | 21/48.2 |
| J. Ordonez-Miranda[18] | T | $VO_2$/Sapphire | Structural | 1.31 | 27/96.5 |
| S. Wang[38] | E and T | PFH-O/PMMA | State | 1.38 | 7/50 |
| W. Kobayashi[26] | E and T | $LaCoO_3$/$La_{0.7}Sr_{0.3}CoO_3$ | - | 1.43 | -233/-173 |
| K. I. Garcia-Garcia[15] | E | Nitinol/Graphite | Structural | 1.47 | 17/177 |
| J. J. Martinez-Flores[19] | E | Gd/Si | Structural | 1.62 | 19/48 |
| R. S. Nakayama[28] | E and T | $Al_{61.5}Cu_{26.5}Fe_{12}$/$Ag_2Te$ | Structural | 1.63 | 27/270 |
| E. Pallecchi[17] | E and T | PNIPAM/PDMS | Structural | 1.96 | 25/45 |
| T. Takeuchi[16] | E and T | $Al_{61.5}Cu_{26.5}Fe_{12}$/$CuGaTe_2$ | - | 2.20 | 27/627 |
| H. Kang[21] | T | Polyethylene/$VO_2$ | Structural | 2.40 | 37/57 |
| A. L. Cotrill[20] | E and T | PFH-O/PNIPAM | State/Structural | 2.60 | 7/44 |
| The present work | E and T | $CaCl_2 \cdot 6H_2O$/paraffin | State | 3.00 | 15/40 |

In the present study, a supercooling-enabled PCTD has been fabricated, experimentally tested and numerically analyzed. Two PCMs with close phase change temperatures are hand-picked, namely, calcium chloride hexahydrate ($CaCl_2 \cdot 6H_2O$) and paraffin. The variable effective thermal conductivities resulted from the involvement of natural convection in the post-melting and supercooled regions are precisely considered in theoretical modeling. The PCTD achieves maximum thermal rectification ratio of 3.0 and outperforms the previous works as can be seen in Table 1. As a comparative implication, supercooling is predicted to be one of the potential interests to sustainably manipulate the heat flux and tune the thermal rectification ratio within easily accessible temperature biases henceforth.

## 2. Design Criteria and Thermo-Physical Properties of Thermal Media

Based on the aforementioned interpretation and classification of PCMs, it is inferred that selecting the right thermal media and establishing the underlying mechanism are instantaneously painstaking as well as crucial. To date, the foremost mechanism of a thermal diode has been realized by varying the thermal conductivity with respect to temperature,[39] as well as through structural modifications[40,41] encompassing dissimilar bulk or thin interfaces. Particularly, thermal media demonstrating temperature-reliant thermal conductivities normally produce two kinds of trends which are termed as positive temperature coefficient (PTC) and negative temperature coefficient (NTC). And the similar trends have also been anticipated for PCMs, thereby suggesting to exploit them for fabrication of phase-change-driven thermal diodes. In principle, PCMs bearing PTC trend demonstrate rise of thermal conductivities upon phase change process, while contrarily, thermal conductivities decrease for PCMs having NTC trend. To fabricate the PCTD with high thermal rectification ratio, the ratio of thermal conductivities of PTC PCM in solid and liquid phases should be small, while the ratio of thermal conductivities of NTC PCM in solid and liquid phases should be large.

By taking the above-mentioned criteria into account, paraffin (PCM A) with a melting temperature $T_{mA}$ of 35 °C, and $CaCl_2 \cdot 6H_2O$ (PCM B) with a melting temperature $T_{mB}$ of 30 °C and a freezing temperature $T_{fB}$ of 7 °C, have been chosen to fabricate the PCTD. Supercooling-induced phase change hysteresis from $T_{mB}$ to $T_{fB}$ has been experimentally determined as elaborated in Section 1 and shown in Figure

S1c of SI. Within the region of phase change hysteresis, the value of $T_{fB}$ allows to achieve spontaneous supercooling release (SSR) and manual supercooling release (MSR). Paraffin bears a large ratio of thermal conductivities in solid ($k_{As}$=0.35 W m$^{-1}$ K$^{-1}$)[42] and liquid ($k_{Al}$=0.16 W m$^{-1}$ K$^{-1}$)[43] phases, which can act as a NTC candidate only if the natural convection in liquid state becomes negligibly small. With the help of theoretical analysis, the height of paraffin terminal was chosen to be 4.8 mm, and natural convection in such a small height is negligible.[44] Furthermore, $CaCl_2 \cdot 6H_2O$ can suitably serve as a PTC material due to its small ratio of thermal conductivities in solid phase ($k_{Bs}$=0.77 W m$^{-1}$ K$^{-1}$), liquid phase ($k_{Bl}$=0.55 W m$^{-1}$ K$^{-1}$) and convection-intensified effective thermal conductivity of liquid phase ($k_{Bl\_eff}$ =4.76 W m$^{-1}$ K$^{-1}$ and 3.15 W m$^{-1}$ K$^{-1}$ at a mean temperature of 35 °C and 20 °C, respectively). The $k_{Bl\_eff}$ quantifies the natural convection of aqueous $CaCl_2$ solution that is further explained in Section 2.3 of SI.

## 3. Theoretical Modeling of the PCTD

Considering a bi-terminal PCTD sandwiched between a heater situated at the bottom and a cooler situated at the top, the heater temperature is set at $T_h$, and the cooler temperature is set at $T_c$. With exception of heat losses from the boundaries, the heat flux through the PCTD is assumed to be unidirectional, flowing from the hot terminal towards cold terminal. A reasonable assumption is made that $k_{As}$, $k_{Al}$, $k_{Bs}$ and $k_{Bl}$ are treated constant because they vary little in the temperature range of 0~40 °C in the present study. However, $k_{Bl\_eff}$ majorly varies with the height of liquid phase. In addition, regarding phase change temperatures with reference to analysis of Cottril et al.,[45] the equivalence of the melting temperatures of two PCMs, i.e., $T_{mA}=T_{mB}$, is supposed to be an ideal condition.

Theoretical modeling has been established through the Fourier's law of heat conduction theory and quantified natural convection in the corresponding forward direction and reverse direction operations. In each operation, the cooling process (performed by reducing the cooler temperature) and heating process (performed by increasing the cooler temperature) were adopted to investigate the effects of supercooling and non-supercooling of $CaCl_2 \cdot 6H_2O$ on the thermal rectification ratio, respectively. A new parameter $T_{fB}$ that serves to critically handle the supercooling effect has been introduced in theoretical modeling. Implementing the successive operations of the PCTD via cooling and heating processes helps configure eighteen thermal modules under various operating conditions, whose theoretical modeling is elaborated in Section 3.1 and 3.2 of SI, respectively. In the heating process, the PCTD modules for PCM A and PCM B can be configured by creating relationships between $T_{mA}$ and $T_{mB}$ through the interface temperature $T_i$, as schematically shown in Table S1 of SI. Likewise, in the cooling process, the PCTD modules for PCM A and PCM B can be developed by relating $T_{mA}$, $T_{mB}$ and $T_{fB}$ via $T_i$ and $T_c$, as schematically shown in Table S2 of SI.

Herein, we explain the method of theoretical analysis by taking the most complicated module as an example, namely Forward-C3, as shown in Table S2 of SI. This module follows the operating condition of $T_{mA}<T_i \leq T_{fB}$ in the forward direction of cooling process, where two PCMs exist in both solid and liquid phases simultaneously, thereby generating the four thermal resistances in total. According to the Fourier's law of heat conduction, the temperature distributions in each junction of the PCTD are considered linear, which are presented as follows:

$$\begin{cases} T_{As}(x) = -\dfrac{T_{mA} - T_c}{L_A - \delta_A}(x - L_B - \delta_A) + T_{mA} \\ T_{Al}(x) = -\dfrac{T_i - T_{mA}}{\delta_A}(x - L_B) + T_i \\ T_{Bs}(x) = -\dfrac{T_{mB} - T_i}{L_B - \delta_B}(x - \delta_B) + T_{mB} \\ T_{Bl}(x) = -\dfrac{T_h - T_{mB}}{\delta_B}x + T_h \end{cases} \quad (2)$$

where $\delta_A$ and $\delta_B$ are the heights of the PCM A and PCM B in liquid phase, respectively.

Since the heat conduction is unidirectional, the heat fluxes through different junctions of the PCTD are equal, being evaluated by the temperature gradient and effective thermal conductivity of the respective PCM, which is expressed as follows:

$$q = \frac{T_{mA} - T_c}{L_A - \delta_A}k_{As} = \frac{T_i - T_{mA}}{\delta_A}k_{Al} = \frac{T_{mB} - T_i}{L_B - \delta_B}k_{Bs} = \frac{T_h - T_{mB}}{\delta_B}k_{Bl\_eff} \quad (3)$$

The heat flux $q$, the interface temperature $T_i$, and the heights $\delta_A$ and $\delta_B$ in liquid phase can be derived from Equations (2) and (3). The mathematical formulae are given as below:

$$\begin{cases} T_i = \dfrac{k_{Bl\_eff}L_A(T_h - T_{mB}) + k_{Bs}L_A(T_{mB} - T_{mA}) - k_{As}L_B(T_{mA} - T_c)}{k_{Al}L_B + k_{Bs}L_A} + T_{mA} \\[6pt] \delta_A = \dfrac{k_{Bl\_eff}L_A(T_h - T_{mB}) + k_{Bs}L_A(T_{mB} - T_{mA}) - k_{As}L_B(T_{mA} - T_c)}{k_{Al}k_{Bl\_eff}(T_h - T_{mB}) + k_{Al}k_{Bs}(T_{mB} - T_{mA}) + k_{As}k_{Bs}(T_{mA} - T_c)} k_{Al} \\[6pt] \delta_B = \dfrac{(k_{Al}L_B + k_{Bs}L_A)(T_h - T_{mB})}{k_{Al}k_{Bl\_eff}(T_h - T_{mB}) + k_{Al}k_{Bs}(T_{mB} - T_{mA}) + k_{As}k_{Bs}(T_{mA} - T_c)} k_{Bl\_eff} \\[6pt] q = \dfrac{k_{Al}k_{Bl\_eff}(T_h - T_{mB}) + k_{Al}k_{Bs}(T_{mB} - T_{mA}) + k_{As}k_{Bs}(T_{mA} - T_c)}{k_{Al}L_B + k_{Bs}L_A} \end{cases} \quad (4)$$

The mathematical formulae of $T_i$, $\delta_A$, $\delta_B$ and $q$ for other thermal modules have been derived in the same manner, as detailed in Section 3 of SI.

In experiments, the forward and reverse heat fluxes through the PCTD encounter a slight decline due to the presence of thermal contact resistance between two terminals and steel blocks. In order to simplify theoretical analysis, thermal contact resistance is neglected in the present study, but the influence of thermal contact resistance is treated as a modifying factor for heat flux and the resulting heat flux $q_c$ is calculated as follows:[46]

$$q_c = q \frac{R_{td}}{R_{td} + R_{c\_td}} \quad (5)$$

Where $q$ is the heat flux without the influence of thermal contact resistance obtained from Equation (4), $R_{td}$ is the thermal resistance of the PCTD being calculated by $R_{td}=(T_h-T_c)/q$, $R_{c\_td}$ is the thermal contact resistance of the PCTD, which is evaluated in Section 2.2 of SI. Using the formulated heat flux $q_c$ in the forward and reverse directions, the thermal rectification ratio in each thermal module can be determined by Equation (1).

## 4. Design Optimization and Assembly of the PCTD Prototype

Design optimization of the PCTD has been executed using dimensionless length $L_{id}$ and dimensionless temperature $T_{id}$, which are defined as follows:[20]

$$L_{id} = \frac{L_A}{L_B} \qquad (6)$$

$$T_{id} = \frac{T_m - T_c}{T_h - T_m} \qquad (7)$$

where $T_m = \frac{T_{mA} + T_{mB}}{2}$ is the average phase change temperature of two PCMs.

Considering the supercooling of aqueous $CaCl_2$ solution, the relationships of both forward and reverse heat fluxes at $L_{id}$ and $T_{id}$ become different in the respective heating and cooling processes. Thus, the thermal rectification ratio of the PCTD can be obtained under four possible conditions, which are shown in Figure 1a to 1d.

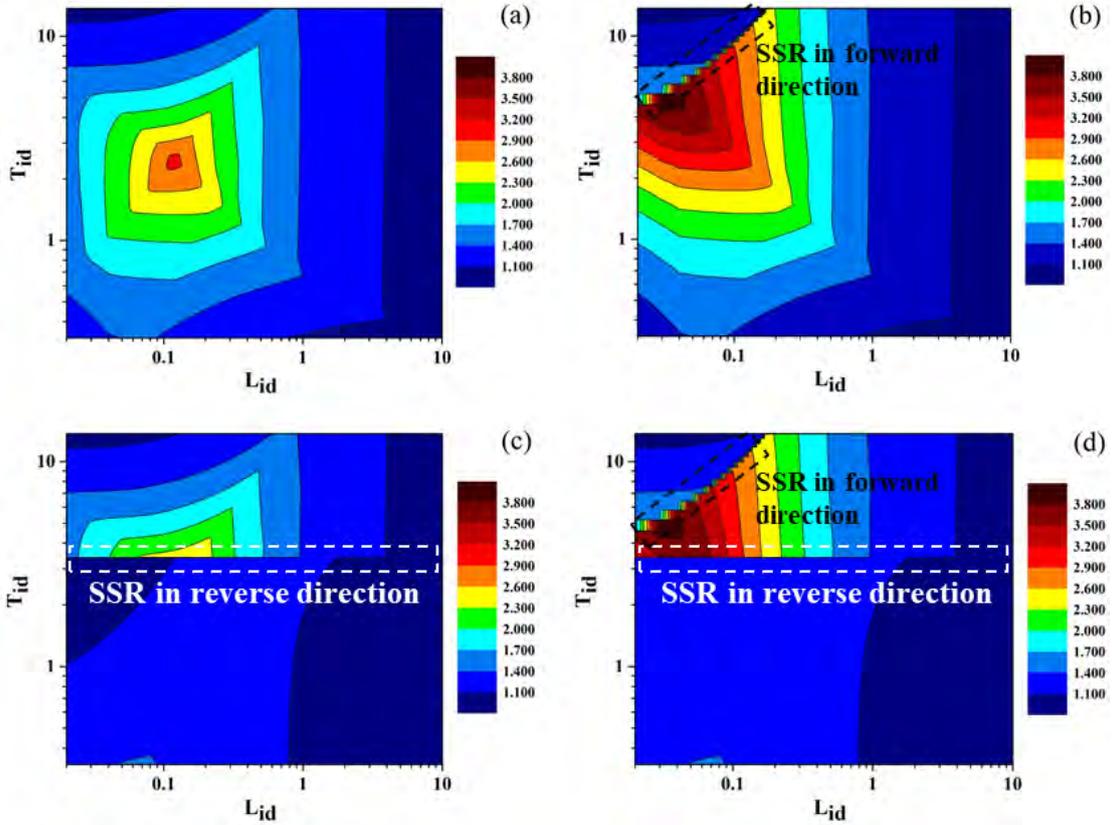

**Figure 1.** Theoretical analysis conducted for design optimization of the PCTD prototype in heating and cooling processes at $L_B$=40 mm and $T_h$=40 °C. The thermal rectification ratios of the PCTD at different $L_{id}$ and $T_{id}$ have been analyzed under four

operating conditions. (a) Both forward and reverse directions operated in heating processes. (b) Forward direction operated in cooling process and reverse direction operated in heating process. (c) Forward direction operated in heating process and reverse direction operated in cooling process. (d) Both forward and reverse directions operated in cooling processes. The maximum thermal rectification ratio of the PCTD using paraffin and $CaCl_2 \cdot 6H_2O$ is 3.0 at $L_{id}$=0.12 and $T_{id}$=2.52.

The evolutions of thermal rectification ratio under different conditions are quite different, and it is quite hard to find an optimum $L_{id}$ and $T_{id}$ which can result in the finest thermal rectification ratio under every condition. Therefore, choosing the maximum thermal rectification ratio of 3.0 at the optimum dimensionless length of 0.12 and optimum dimensionless temperature of 2.52 optimizes the acceptable design and fabrication guideline. Meanwhile, the theoretical results also indicate that the thermal rectification ratio decreases immediately when the $L_{id}$ and $T_{id}$ deviate from the optimum values under heating process, as shown in Figure 1a, while the thermal rectification ratio can retain high value for a wide range of $T_{id}$ when the forward direction is under cooling process, as shown in Figure 1b, which occurs due to supercooling effect. In addition, the contours demonstrate two regions of sudden variation. The weird shapes of contours occurs due to the SSR of aqueous $CaCl_2$ solution. In the reverse direction, when SSR occurs, thermal rectification ratio shows a sudden rise, as shown by the white dashed boxes in Figure 1c and 1d. In the forward direction, when SSR takes place, thermal rectification ratio drops suddenly which depicts sawtooth shape of contours, as shown by the black dashed boxes in Figure 1b and 1d. Subsequently, the PCTD prototype has been fabricated under the guideline of the theoretical analysis, as schematized in Figure 2, depicting the general assembly in the forward and reverse directions.

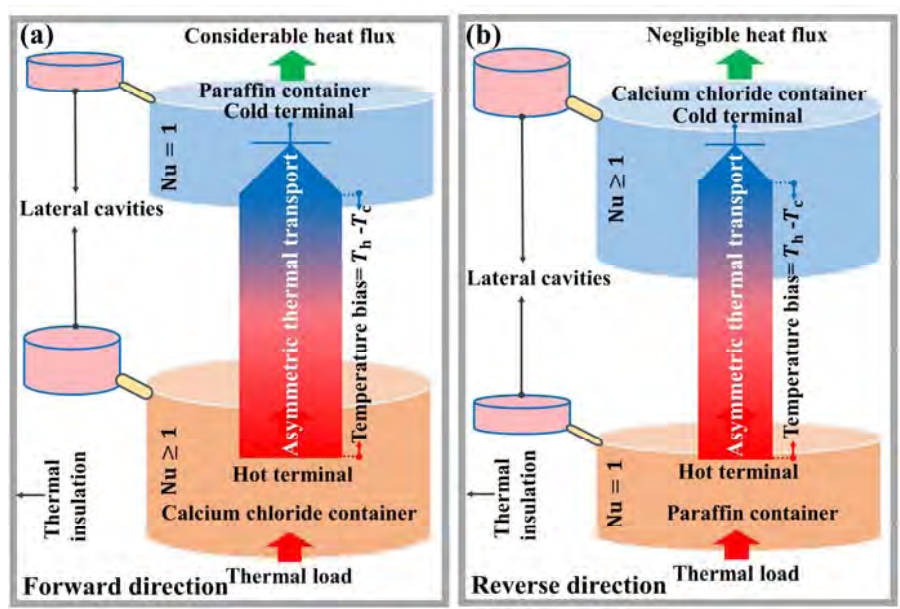

**Figure 2.** Schematic illustration (not to scale), describing the general assembly and working principle of the PCTD. A considerable temperature bias is developed across thermal module under asymmetric thermal transport mechanism driven by physical states of thermal media, leading to a large heat flux in the forward direction, while a negligible heat flux in the reverse direction. Scaled diagram of the prototype can be seen in Figure S5 of SI.

The PCTD module consists of two cylindrical containers: a paraffin terminal with a length of 4.8 mm and a $CaCl_2$ terminal with a length of 40 mm, equaling the optimum terminal lengths ratio of 0.12. The bi-terminals with the same diameter of 50 mm are joined together using conductive thermal grease. Novel lateral cavities have been introduced to avoid the volumetric expansion/contraction of PCMs during long-term operations. In the forward direction, $CaCl_2$ terminal was located adjacent to the bottom heater; while paraffin terminal resided close to the top cooler. In the reverse direction, the positions of both terminals are opposite to those in the forward direction. The further fabrication detail and experimental procedure are mentioned in Section 5 of SI.

## 5. Experimental and Theoretical Results

### 5.1. Heat Flux in the Forward Direction Operation

A series of experimental and theoretical investigations have been systematically conducted both for the forward direction and reverse direction. In the forward direction, variation of heat fluxes versus implemented temperature bias of 10~40 °C is shown in Figure 3a, and the physical states of thermal media at the relevant ranges of cooler temperature have been projected to interpret the underlying mechanism of heat transfer, as shown in Figure 3b.

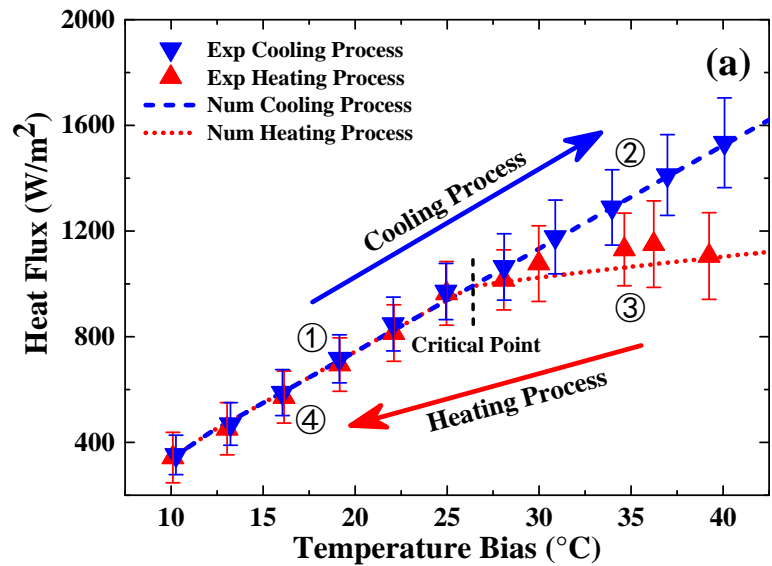

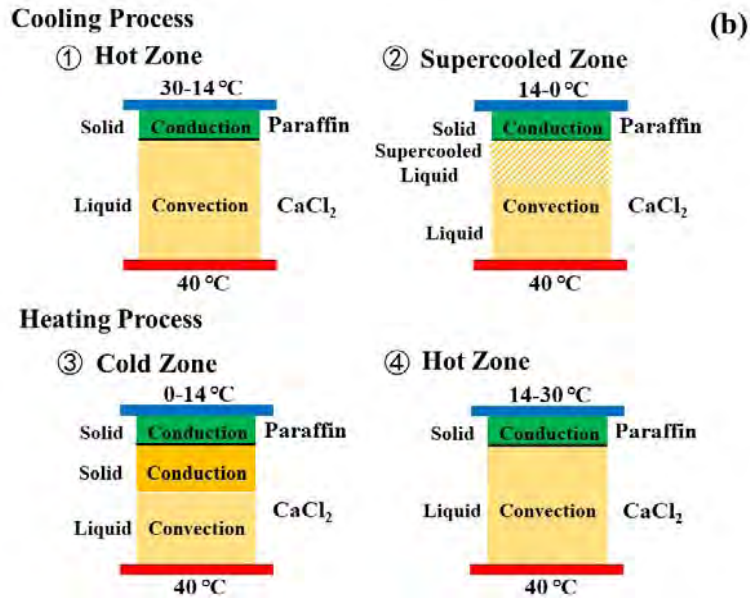

**Figure 3.** Experimental (Exp) and numerical (Num) heat fluxes through the PCTD as a function of temperature bias in the forward direction. (a) Steady state freeze-thaw approach leading to four thermal zones in the temperature bias of 10~40 °C. The temperature of the heater is at 40 °C, while the temperature of the cooler varies in the temperature range of 0~30 °C. (b) Inferred from theoretical results, the schematics show the physical states of thermal media within various ranges of the cooler temperature in the forward direction. The bottom red bar is the heater operated at a constant temperature of 40 °C. The top blue bar exhibits the cooler whose temperature has been varied from 30 °C to 0 °C during the cooling process, and oppositely for the heating process. The cooler temperature range from 30 °C to 0 °C facilitates the corresponding temperature bias from 10 °C to 40 °C, and vice versa.

The experimental and theoretical results are shown in a great agreement in Figure 3a. In addition, four prominent thermal zones have been noticed and symmetrically categorized into the cooling process: hot zone and supercooled zone, and the heating process: cold zone and hot zone. Furthermore, they are distinguished with respect to temperature bias of 26 °C as predicted from theoretical analysis, which is termed as a critical point henceforth.

In the cooling process, the physical state of paraffin terminal remains solid due to its melting temperature higher than the maximum applied temperature of the cooler, so the emerged thermal zones are predominantly resulted from the alternate physical states of $CaCl_2$ terminal. For example, hot zone entails the hot liquidus state at the cooler temperature range of 30~14 °C (the relevant temperature bias is 10~26 °C), as shown in Figure 3b-①; And supercooled zone includes supercooled state along with reserved-liquidus state at the cooler temperature range of 14~0 °C (the relevant temperature bias is 26~40 °C), as schematized in Figure 3b-②. Shown as ① in Figure 3a (temperature bias of 10~26 °C), the hot zone appearing below the critical point is characterized with an almost linear trend of the forward direction heat flux. This behavior is well-elaborated in accordance with achievable physical states of thermal media, as can be seen in Figure 3b-① (cooler temperature range of 30~14 °C). Herein, $CaCl_2$ terminal maintains the temperature sufficiently higher than its melting temperature (30 °C), whose effective thermal conductivity is 4.76 W m$^{-1}$ K$^{-1}$ (experimentally measured at 35 °C), while the paraffin terminal is in solid state with thermal conductivity of 0.35 W m$^{-1}$ K$^{-1}$. As long as the temperature of the cooler is reduced to 14 °C (the relevant temperature bias is 26 °C), the temperature of upper part of $CaCl_2$ terminal is anticipated to approach its phase change temperature according to the numerical results. Along with reducing the cooler temperature continuously from 14 °C to 0 °C (the relevant temperature bias from 26 °C to 40 °C), supercooled zone is approached, as shown by ② in Figure 3a, depicting the gradual conversion of hot liquidus state of $CaCl_2$ terminal into supercooled state whose thermal conductivity is 3.15 W m$^{-1}$ K$^{-1}$ (experimentally measured at 20 °C). As thermal conductivity of supercooled aqueous $CaCl_2$ solution is still high enough to maintain the asymmetric thermal conductance path, the supercooled zone keeps fostering the natural convection until the entire aqueous $CaCl_2$ solution becomes solid on supercooling elimination. In a word, in the cooling process of forward direction operation, the bottom liquidus, the middle supercooled and the top solidus physical states altogether proliferate the asymmetric thermal conductance on behalf of high thermal conductivities across the entire thermal module, yielding preferential heat transfer through well-defined natural convection and conduction.

In the heating process, sequential cold zone and hot zone come into being, consisting of well-preserved solid state of paraffin terminal, and multiple physical states of $CaCl_2$ terminal. For examples, fractional liquidus state and solidus state at the cooler temperature range of 0~14 °C (the relevant temperature bias is 40~26 °C), as shown by ③ in Figure 3b; and the complete liquidus state at the cooler temperature range of 14~30 °C (the relevant temperature bias is 26~10 °C), as shown by ④ in Figure 3b. The cold zone is marked with gradual heat flux downfall as temperature bias is decreased as depicted by ③ in Figure 3a. And alongside that, fractional solidus part of $CaCl_2$ terminal (Figure 3b-③) keeps gradually diminishing until the hot zone (Figure 3a-④) appears at cooler temperature of 14~30 °C (the relevant temperature bias is 26~10 °C). Furthermore, with decreasing temperature bias across the terminals, heat flux tends to fall down because of the reduced temperature gradient and accordingly reaches the minimum magnitude at a temperature bias of 10 °C. In addition, it is worth mentioning that the heat flux determined by numerical analysis agrees well with the experimental results in both the cooling and heating processes, as depicted in Figure 3a, therefore validating the reliability of the adopted theoretical model.

## 5.2. Heat Flux in the Reverse Direction Operation

In the reverse direction, the heat fluxes through the PCTD as a function of temperature bias are shown in Figure 4a, and the physical states of thermal media at the corresponding ranges of cooler temperature are shown in Figure 4b.

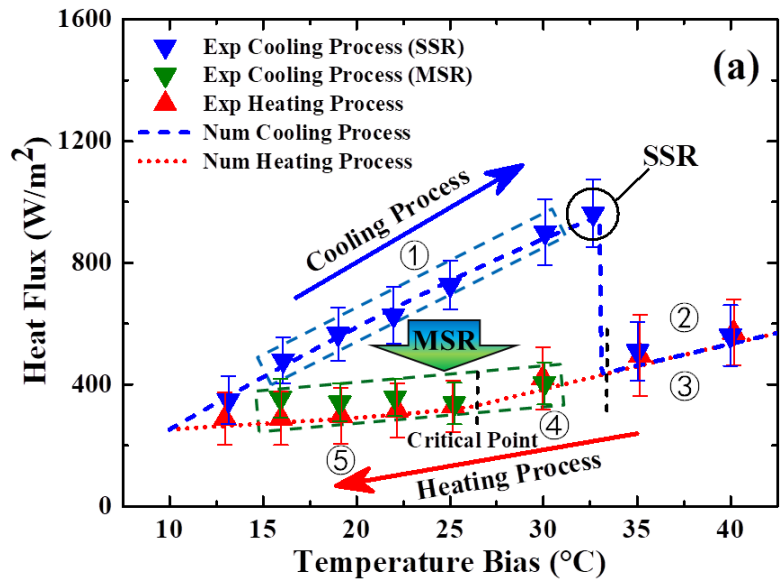

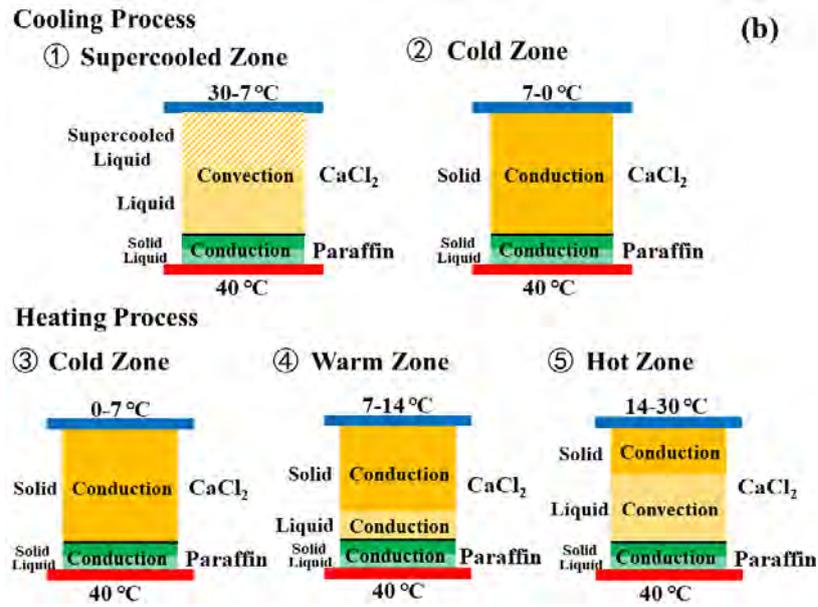

**Figure 4.** Experimental (Exp) and numerical (Num) heat fluxes through the PCTD as a function of temperature bias in the reverse direction. (a) Steady state freeze-thaw approach leading to five thermal zones in the temperature bias of 10~40 °C. The temperature of the heater is at 40 °C, while the temperature of the cooler varies in the temperature range of 0~30 °C. (b) Inferred from theoretical results, the schematics show the physical states of the PCTD within various ranges of the cooler temperature in the reverse direction. The bottom red bar is the heater operated at a constant temperature of 40 °C. The top blue bar exhibits the cooler whose temperature has been varied from 30 °C to 0 °C in the cooling process, and oppositely for the heating process. The cooler temperature range of 30 °C to 0 °C facilitates the corresponding temperature bias range of 10 °C to 40 °C, and vice versa.

Similar to the forward direction operation, the working of the PCTD in the reverse direction has also been split into the cooling and heating processes. Corresponding to the cooling process, the supercooling effect has been particularly investigated in order to find out the performance limits through which the tunability range can thus be identified.

The appearance of the critical point, that is 26 °C, is alike to that in the forward direction due to the same optimum length ratio of $CaCl_2$ and paraffin terminals, but the origination of the asymmetric thermal transport is obviously opposite. Having the constant temperature of heater (40 °C) higher than the melting temperature of paraffin (35 °C), the paraffin terminal maintains both solid and liquid states throughout the reverse direction operation in either cooling or heating process, as shown in Figure 4b. But $CaCl_2$ terminal undergoes multiple physical states at different cooler temperatures, as shown in Figure 4b. In the cooling process, the supercooled zone initially begins at the cooler temperature of 30 °C (the relevant temperature bias is 10 °C). As the temperature of the cooler extends down to nearly 7 °C (the relevant temperature bias is around 33 °C), the aqueous $CaCl_2$ solution tends to reserve supercooled state instead of adopting solid state, as shown by ① in Figure 4b. Since thermal conductivity of supercooled aqueous $CaCl_2$ solution (3.15 W m$^{-1}$ K$^{-1}$ experimentally measured at 20 °C) is quite large, the rising profile of heat flux has evidently been maintained, as represented by ① in Figure 4a. With regard to the cooler temperature of approximately 7 °C (the relevant temperature bias is 33 °C), the freezing point of aqueous $CaCl_2$ solution is approached at which SSR occurs, explaining the sudden drop-down trend in heat flux curve. Henceforth, the cold zone takes place as the temperature of the cooler goes down to 0 °C (the relevant temperature bias is 40 °C), as shown by ② in Figure 4a, at which solid state of $CaCl_2$ terminal (Figure 4b-②) is attained.

Unlike the heat flux profile in the cooling process, gradual reduction of heat flux has been witnessed in the heating process, as can be seen in ③-cold zone, ④-warm zone and ⑤-hot zone of Figure 4a. The physical states of $CaCl_2$ terminal demonstrate gradual interconversion within the entire cooler temperature range of 0 °C to 30 °C (the relevant temperature bias is from 40 °C to 10 °C), as seen from the schematics ③, ④ and ⑤ of Figure 4b. Such a trend occurs because of decreasing temperature bias, and particularly the heat flux curve bears a distinct behavior at the critical point where the melting of $CaCl_2 \cdot 6H_2O$ is expected with a part of liquidus state.

Based on the above results, it is postulated that the heat flux in the cooling process during supercooled zone is relatively large even in the reverse direction which seems to be a drawback of the PCTD. However, the supercooling effect can be conveniently controlled at any desired temperature within supercooled zone. In order to realize it, MSR is implemented by crystal seeding technique. The entire procedure is explained in Section 6 of SI and schematically shown in Figure S7 of SI. As soon as the solid crystals of $CaCl_2 \cdot 6H_2O$ are seeded into the supercooled liquid, aqueous solution tends to achieve the solid state, as shown by ⑤ of Figure 4b, during which the heat flux is limited to minimum magnitudes and the evidence is clear from the hot zone (⑤) of Figure 4a (green downward triangular symbols encased in a dotted green rectangle). As a precaution in the reverse direction in the cooling process, supercooling should be preferentially avoided to limit the heat flux. Moreover, theoretical analysis finds a reliable validation through a good agreement with experimental heat flux in both the cooling and heating processes.

## 5.3. Thermal Rectification Ratio and Tunability

By using the heat fluxes of the forward direction and reverse direction operations, as shown in Figure 3a and Figure 4a, respectively, the corresponding thermal rectification ratios are calculated and numerically analyzed, as shown in Figure 5.

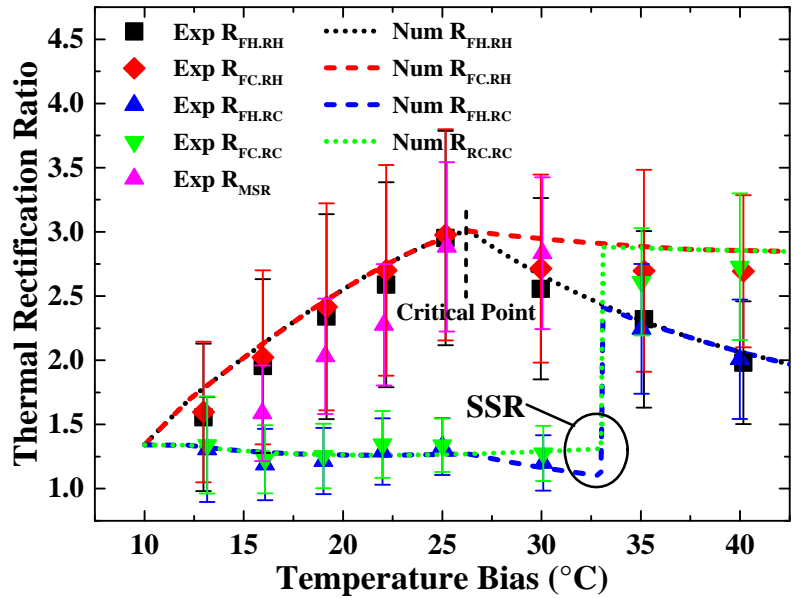

**Figure 5.** Experimental (Exp) and numerical (Num) thermal rectification ratios ($R$) obtained by using Equation (1) through various combinations of heat fluxes, such as $R_{FH.RH}$: forward heating (FH) to reverse heating (RH); $R_{FC.RH}$: forward cooling (FC) to reverse heating (RH); $R_{MSR}$: forward cooling to reverse cooling with manual supercooling release (MSR); $R_{FH.RC}$: forward heating (FH) to reverse cooling (RC); $R_{FC.RC}$: forward cooling (FC) to reverse cooling (RC).

Two major categories of thermal rectification ratios can be realized: the maximum thermal rectification ratios abbreviated as $R_{FH.RH}$, $R_{FC.RH}$ and $R_{MSR}$; the minimum thermal rectification ratios abbreviated as $R_{FH.RC}$ and $R_{FC.RC}$ (the corresponding definitions are enlisted in the caption of Figure 5).

It is eventually exposed that the physical states steadily contribute to asymmetric thermal conductance, and consequently, maximum thermal rectification of 3.0 has been achieved. In addition, the sustainability of thermal rectification ratio spreading over a wide temperature bias is no more questionable, as evidenced from red curve in Figure 5. Contrary to the conventional thermal transport mechanism lacking of supercooling effect, it is hereby uncovered that supercooled zone characterized with relatively high effective thermal conductivity acts as a leverage, ensuring a greater drive for heat flux manipulations. For example, consider $R_{FH.RH}$ (black curve in Figure 5) where supercooled zone is absent, it typically elucidates the convectional thermal performance. Symbolically, $R_{FH.RH}$ rises non-linearly within temperature bias of 10~26 °C due to overall high thermal conductivity of thermal media in the forward direction (aqueous $CaCl_2$ solution: 4.76 W m$^{-1}$ K$^{-1}$ at 35 °C, Paraffin: 0.35 W m$^{-1}$ K$^{-1}$) and low thermal conductivity of thermal media in reverse direction. Soon above the critical point, $R_{FH.RH}$ encounters an immediate downfall within temperature bias of 26~40 °C. The reason is attributed to rapidly emerging strong thermal barrier owing to overall reduced thermal conductivity of thermal media in the forward direction of heating process (FH), exemplifying the unstable thermal performance. Besides, within temperature bias of 10~26 °C, $R_{FC.RH}$ (red curve) also comes up with a rising trend as $R_{FH.RH}$ does, however, the trend for the former preserves the steadiness even above the critical point. The emergence of sustainable $R_{FC.RH}$ is supported by the supercooling effect of aqueous $CaCl_2$ solution in the forward direction of cooling process (FC) during which heat is transported by extended natural convection. In addition, the observed supercooling-enabled sustainability is advantageous to tune the thermal diode in a wide temperature bias of 10~33 °C. As represented by magenta color points in Figure 5, $R_{MSR}$ has a growing trend but tends to be marginally below as compared with $R_{FC.RH}$ and $R_{FH.RH}$. This is most likely ascribed to the non-uniformity introduced by seeding solid crystals of $CaCl_2 \cdot 6H_2O$, leading to the solidification of aqueous $CaCl_2$ solution, and consequently lowering down the natural convection. Indeed, the seeded crystals create nucleation sites and compel the supercooled zone into solidus zone whose thermal conductivity is much smaller as compared with the former zone. Therefore, thermal rectification ratio increases comparatively slow, but opens up a gate to tune the thermal diode by eliminating the supercooling effect.

By operating thermal diode in the cooling process, especially in the reverse direction, two thermal rectification ratios have been generated denoted as $R_{FH.RC}$ and $R_{FC.RC}$. Both thermal rectification ratios are precisely predicted in view of the SSR, i.e., auto-conversion of supercooled aqueous $CaCl_2$ solution into fully solid state at temperature of approximately 7 °C. Since the entire aqueous $CaCl_2$ solution does not change into solid state instantaneously, the part-wise change of physical state of thermal media becomes favorable to provide an approximately horizontal trend of thermal rectification ratio along the temperature bias of 10~33 °C, as shown by green and blue curves in Figure 5. As the temperature bias of 33 °C is reached, solid state with low thermal conductivity appears, thereby bringing about a sudden rise in thermal rectification ratio due to the low heat flux in the reverse direction. However, the abnormal trend during SSR help recommend that thermal diode should not be operated in the cooling process of reverse direction.

## 6. Governing Theory and Mechanism of the PCTD

The present study instigates to utilize the unexplored thermo-physical properties of PCMs, as well as contributes to establish the fundamental governing theory necessary to understand the dynamics and inter-dependent parameters of a PCTD. The findings depict that not only the materials demonstrating thermal conductivity variances and length ratio of terminals are vital, but also other thermo-physical parameters such as variable intensity of natural convection, phase change hysteresis and melting temperature difference are worthy of considerations. Moreover, it is postulated that a phase change thermal medium can either undermine the maximum thermal rectification ratio or strengthen it on behalf of a gap between the state-specific thermal conductivities. Precisely, with advantage of a large gap between the (effective) thermal conductivities in solid and liquid phases, a remarkable thermal rectification ratio can be achieved. With in-depth theoretical and experimental analyses, we thus propose the overall governing mechanism, namely phase change-driven asymmetric thermal transport, as schematically shown in Figure 6.

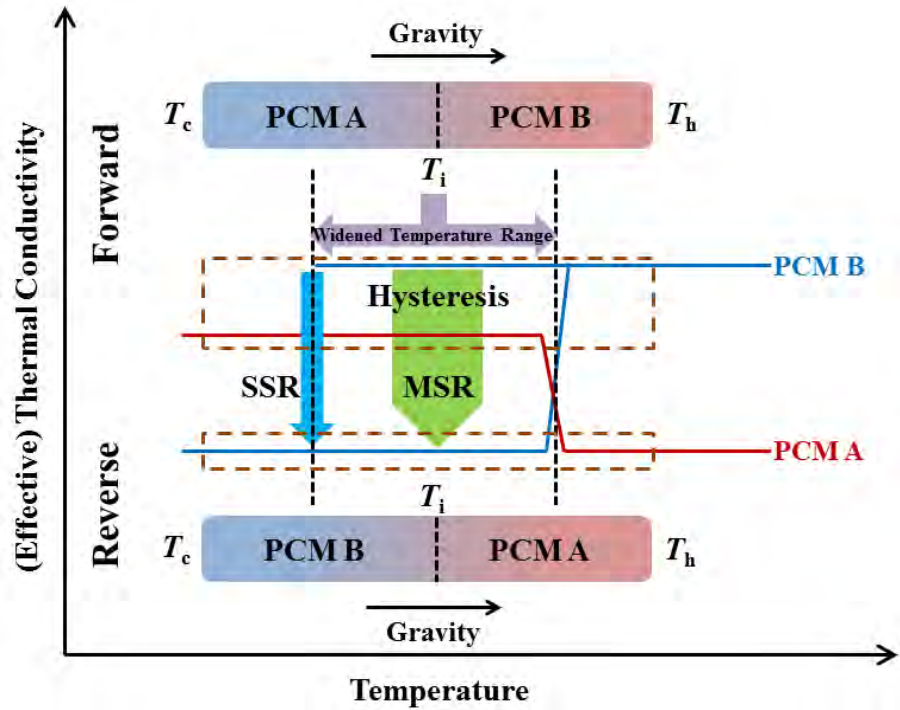

**Figure 6.** Phase change-driven asymmetric thermal transport: Governing mechanism of as-fabricated PCTD (not to scale) referring to the best performance in terms of maximum and sustainable thermal rectification ratio. The paraffin terminal (PCM A) and $CaCl_2$ terminal (PCM B) mutually exhibit NTC (red curve) and PTC (blue curve) trends, respectively, and aqueous $CaCl_2$ solution shows phase change hysteresis due to supercooling effect. State-specific thermal conductivities build up asymmetric thermal transport in the forward and reverse directions (dashed rectangles).

Although temperature-dependent physical-state variations give birth to variable thermal conductivities (i.e., NTC and PTC trends), enhancement in thermal rectification ratio also depends on the intrinsic thermo-physical nature of thermal media. For example, paraffin physically undergoes phase change but is capable to keep thermo-physical properties constant in either states. Similarly, phase change hysteresis, as displayed in Figure 6, is a promising feature of aqueous $CaCl_2$ solution which has been found to be responsible for a durable as well as a tunable function of the PCTD.

As illustrated in Figure 6, both PCM terminals lie in the high thermal conductive region (upper dashed rectangle) in the forward direction, while the reverse direction induces high thermal resistive region (lower dashed rectangle). Interface temperature ($T_i$) is a decisive factor whose monitoring makes the phase change hysteresis functional, and also imparts vigorous performance to the PCTD. Based on $T_i$, the proposed mechanism can be generalized to explain the performance of supercooling-enabled (advanced) and non-supercooling-enabled (traditional) PCTDs. For the latter case, once $T_i$ reaches the phase change temperature in the forward direction, (effective) thermal conductivities of two PCMs becomes high, and in response, thermal rectification ratio also achieves the maximum value; however, this value is temporal and the PCTD performance soon faces a rigorous decline (e.g., the black curve in Figure 5). This is ascribed to the imbalance of (effective) thermal conductivities created because of progressive state-change of thermal media. Unlike that deteriorated performance, supercooling-enabled PCTD works quite efficiently by making use of phase change hysteresis where a sustainable thermal rectification ratio (e.g., the red curve in Figure 5) is attributed to $T_i$ monitoring within a widened temperature range. For this, the phenomenological theory is based on the strong convective junction (supercooled and hot liquidus states) developed within domain of aqueous $CaCl_2$ solution. Briefly, (effective) thermal conductivities of PCMs overall become high (the upper dashed rectangle), and consequently an extensive heat flux lasts longer until $T_i$ becomes equal to the freezing point.

On the other hand, as shown in Figure 6, MSR provides a reliable control over temperature during phase change hysteresis, rendering the PCTD as a tunable thermal device. Since tunability is a factor that allows thermal diode to dynamically change the thermal rectification ratio, the possible applications to the transient processes, such as thermal logic circuitry, thermal computing and information processing, etc., can be potentially enabled.

Moreover, geometrical arrangement of physical module also helps elaborate the underlying mechanism of directional heat flows. To realize the concept, we introduce the simplest vertical model consisting of phase change media sandwiched between a bottom heater and a top cooler. For the existence of natural convection, the development of temperature gradient in upward direction is imperative, and further natural convection can be adequately strengthened or extremely weakened by controlling the size of PCM chambers. Since aqueous $CaCl_2$ solution sustains natural convection in a widened temperature range due to phase change hysteresis, the size of $CaCl_2$ chamber is intentionally kept large. While, the smaller size of paraffin chamber assists in suppressing the natural convection and results in only conduction-dominant heat transfer. Consequently, the entire arrangement brings about asymmetrical thermal conductance, providing a considerable thermal rectification ratio.

In summary, thermal performance of a thermal diode should be optimized by manipulating the state-specific heat fluxes so that the best combination of the forward and reverse heat fluxes can be proposed henceforth, leading to a superior thermal rectification ratio.

## 7. Conclusion

A novel supercooling-enabled PCTD is presented, setting forth the theoretical design guidelines and routes to optimize its performance. Theoretical hypothesis leads to fabricate a prototype thermal diode with an optimum length ratio of 0.12, consisting of NTC paraffin terminal and convection-intensified PTC $CaCl_2$ terminal. Findings reveal that considerable heat flux in the forward direction operation is driven by convection-intensified liquid states, while conduction prevails in consecutive solid (or limited liquid) state known for high thermal resistance in reverse direction. More significantly, introduction of $T_{fB}$ in theoretical modeling serves as a novel parameter which provides manifold benefits, including a widened temperature range, extensive convection empowered by phase change hysteresis due to supercooling effect, sustainable thermal rectification and tunable thermal performance at any temperature between 7~30 °C. Consequently, a giant thermal rectification ratio of 3.0 is obtained. Cumulatively, the conclusive experimental and theoretical evidences help understand and establish the novel governing mechanism combining the state-specific asymmetric thermal conductance with phase change hysteresis of $CaCl_2$ terminal.

## 8. Experimental Section

*Materials and characterization*: Paraffin (Ruhr Tech, OP35E) and calcium chloride anhydrous ($CaCl_2$) (SCR, AR, >96%) were purchased and employed as-received.

CaCl$_2$ was dissolved in deionized water with mass ratio of 111:108 to obtain aqueous CaCl$_2$ solution, while the frozen state of aqueous CaCl$_2$ solution has been termed as CaCl$_2$·6H$_2$O. Phase change temperatures of both materials were determined through differential scanning calorimeter (Perkin Elmer, DSC 8000) operated at a heating rate of 5 °C min$^{-1}$ under inert atmosphere of N$_2$. Furthermore, thermal conductivity of CaCl$_2$·6H$_2$O was determined through a steady-state method, as elaborated in Section 2 of SI. Likewise, supercooling degree of CaCl$_2$·6H$_2$O was measured with help of a customized experiment. A glass beaker filled with 30 ml aqueous CaCl$_2$ solution was heated up to 50 °C. Then, the heated aqueous CaCl$_2$ solution was immersed in chilled ethylene glycol-water mixture in a thermostatic bath (DC-2020, CNSHP) maintained at -10 °C. Having a resistive temperature detector (RTD Pt-100) inserted at the center of aqueous CaCl$_2$ solution, the transient temperature history was measured through data acquisition system (Keithley 2700) and subsequently recorded as a function of time, as graphically shown in Figure S1c of SI.

*Prototype fabrication and experimental methodology*: As-proposed theoretical model of the PCTD has been physically crafted into a prototype consisting of alternative paraffin (length: 4.8 mm) and CaCl$_2$ (length: 40 mm) terminals. The process and instrumentation diagram of experimental set-up is elaborated in Section 5 of SI and schematically shown in Figure S5 (a, b and c). Temperature gradient across the PCTD was measured by adjusting the prototype into test section of a customized vertical set-up, as shown in Figure S5 (a-right). The test section was actually a sandwiched chamber clamped with outward-extended two steel blocks at the top and bottom.

Furthermore, the steel blocks were connected with a bottom heater and a top cooler, facilitating constant temperature hot water and cold ethylene glycol, respectively. Throughout the experimental runs, temperature of the bottom heater was kept constant at 40 °C, while the temperature of top cooler was varied from 0 °C to 30 °C in the heating process, and from 30 °C to 0 °C in the cooling process. Therefore, the temperature bias was monitored in the range of 10~40 °C. To measure the temperature data across the PCTD, four equidistant resistive temperature sensors (RTD Pt-100) were centrally inserted into the top and bottom steel block. After the PCTD achieved steady-state, the temperature history was recorded using data acquisition system (Keithley 2700) with subsequent heat flux calculations according to the Fourier's law of heat conduction.

**Supporting Information**

Supporting Information is available from the Wiley Online Library or from the author.

**Conflict of Interest**

The authors declare no conflict of interest.

**Acknowledgements**

This research is supported by the National Natural Science Foundation of China under the contract No. 51676122. A part of the characterizations are conducted in AEMD of Shanghai Jiao Tong University.

# Supporting Information

**Supercooling-enabled giant and tunable thermal rectification ratio of a phase change thermal diode**

*Zhaonan Meng, Raza Gulfam, Peng Zhang\*, Fei Ma*

Institute of Refrigeration and Cryogenics, Shanghai Jiao Tong University, No. 800 Dongchuan Road, Shanghai 200240, China

E-mail: zhangp@sjtu.edu.cn

1. Determination of phase change temperature of thermal media

The phase change temperatures of paraffin and $CaCl_2 \cdot 6H_2O$ were determined through differential scanning calorimeter (DSC). Under heating rate of 5 °C min$^{-1}$, paraffin and $CaCl_2 \cdot 6H_2O$ were tested within temperaure range of 0 °C to 50 °C and -10 °C to 50 °C, respectively. According to the DSC curves, as shown in Figure S1(a and b), the phase change temperature of paraffin is $T_{mA}$=35 °C, while that of $CaCl_2 \cdot 6H_2O$ is $T_{mB}$=30 °C. Meanwhile, the degree of supercooling of $CaCl_2$ solution was also measured experimentally. A small amount of $CaCl_2 \cdot 6H_2O$ (about 30 ml) was put in a beaker and subsequently heated in a water bath at 50 °C. When the $CaCl_2 \cdot 6H_2O$ was melted into aqueous $CaCl_2$ solution completely, it was put into a glycol bath with temperature of -10 °C. A resistive temperature detector (RTD) was immersed in aqueous $CaCl_2$ solution to measure the freezing temperature $T_{fB}$. According to the cooling curve shown in Figure S1c, the $T_{fB}$ aqueous $CaCl_2$ solution is 7 °C. Therefore, the corresponding phase change hysteresis spans from $T_{mB}$ of 30 °C to $T_{fB}$ of 7 °C.

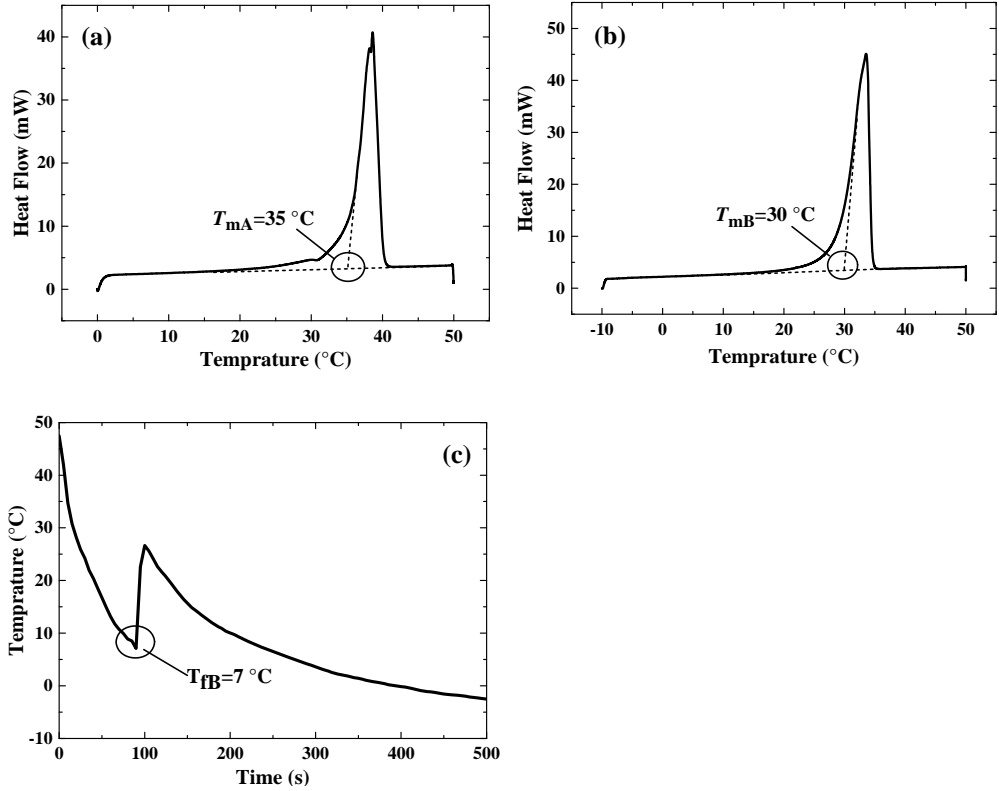

Figure S1. (a) Melting temperature of paraffin measured by DSC from 0 °C to 50 °C. (b) Melting temperature of CaCl$_2$·6H$_2$O measured by DSC from -10 °C to 50 °C. (c) Freezing temperature of aqueous CaCl$_2$ solution from 50 °C to below 0 °C.

2. Measurement of thermo-physical properties of thermal media

2.1 Thermal conductivities of CaCl$_2$·6H$_2$O in solid and liquid phases

Thermal conductivities influence the performance of the PCTD, necessitating the precise determination. Thermal conductivities of paraffin in solid and liquid phases were adopted as $k_{As}$ =0.35 W m$^{-1}$ K$^{-1}$[S1] and $k_{Al}$= 0.16 W m$^{-1}$ K$^{-1}$[S2], respectively, while thermal conductivities of CaCl$_2$·6H$_2$O in solid phase $k_{Bs}$ and liquid phase $k_{Bl}$ were measured through the steady-state method that was conducted in a custom-built experimental device[S3]; and the schematic of experimental set-up is shown in Figure S2. The set-up contained a sample with a diameter of 50 mm enclosed by the bottom and top steel blocks which were further affixed with a heater and a cooler, respectively. The function of steel blocks was to help measure the unidirectional heat flux across two ends of the sample. Meanwhile, the entire experimental device was covered by thermal insulation material to avoid the heat loss. To measure the temperature gradient d$T$/d$x$ and obtain heat flux $q_s$ by means of Equation S1, eight RTDs were inserted in both steel blocks along the central axis at different heights.

$$q_s = -k_{st}\frac{dT}{dx} \tag{S1}$$

Where, $k_{st}$ is thermal conductivity of the steel blocks, which is temperature-dependent and can be estimated through the following equation:

$$k_{st} = 0.0525T_{st} + 11.05 \tag{S2}$$

Where, $T_{st}$ is the mean temperature of the steel blocks.

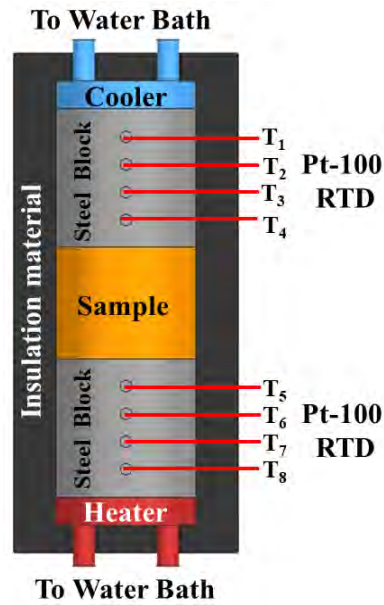

Figure S2. Experimental set-up for thermal conductivity measurement.

The temperatures at the top and bottom of the sample can be determined by extrapolating the temperature distribution, and the total thermal resistance $R_t$ can be determined by the following equation:

$$R_t = R_s + R_{c\_s} = \frac{\Delta T_s}{q_s} \tag{S3}$$

Where, $\Delta T_s$ is the temperature difference between two ends of the sample, $R_s$ denotes thermal resistance of the sample. $R_{c\_s}$ represents thermal contact resistance of the two interfaces between the sample and two steel blocks. In the experiment, thermal grease was pasted between all interfaces of the sample and steel blocks, therefore thermal contact resistances for different interfaces deemed to be very small. To obtain thermal conductivity $k_s$ of the sample and value of $R_{c\_s}$, two thermal resistances $R_{t1}$ and $R_{t2}$ of the samples with different thicknesses of $h_1$ and $h_2$ were measured. Therefore, $k_s$ and $R_{c\_s}$ can be calculated as follows:

$$k_s = \frac{h_1 - h_2}{R_{t1} - R_{t2}} \tag{S4}$$

$$R_{c\_s} = R_{t1} - \frac{h_1}{k_s} = R_{t2} - \frac{h_2}{k_s} \tag{S5}$$

In the present study, two samples of $CaCl_2 \cdot 6H_2O$ were tested with thickness of 20 mm and 40 mm. Thermal resistance in solid and liquid phases was measured by the custom-built experimental set-up, as shown in Figure S2. Most significantly, thermal resistance of liquid can be measured accurately only if the natural convection is avoided. To do so, the heater was switched from the bottom to the top which resulted in approximately no convection in the downward direction, allowing to determine thermal conductivity in the most precise way. As evaluated by Equation S4 and S5, thermal conductivity of $CaCl_2 \cdot 6H_2O$ in solid phase $k_{Bs}$ is 0.77 W m$^{-1}$ K$^{-1}$ at 20 °C and in liquid phase $k_{Bl}$ is 0.55 W m$^{-1}$ K$^{-1}$ at 35 °C, and thermal contact resistance $R_{c\_s}$ of the sample is 0.003 K m$^2$ W$^{-1}$.

2.2 Thermal contact resistance of the PCTD

The influence of the thermal contact resistance should be necessarily taken into consideration. The value of the thermal contact resistance in Section 2.1 of SI is 0.003 K m$^2$ W$^{-1}$ that includes two interfaces between the single sample and two steel blocks. However, when the single sample is replaced by thermal diode, an additional thermal contact resistance at the interface of two clamped PCMs should be considered whose value equals to that of interface between the single sample and steel block. As a consequence, the thermal contact resistance $R_{c\_td}$ of thermal diode becomes 1.5 times larger than the $R_{c\_s}$ of the single sample, thus $R_{c\_td}$=0.0045 K m$^2$ W$^{-1}$ has been cautiously used while carrying out theoretical analysis of the PCTD.

2.3 Effective thermal conductivity of aqueous $CaCl_2$ solution under natural convection

In the mathematical model, the natural convection during hot liquidus states and supercooled states, particulalry of of aqueous $CaCl_2$ solution, has been quantified in terms of effective thermal conductivity. It can be formulated as follows:

$$k_{Bl\_eff} = Nu k_{Bl} \tag{S6}$$

Where, Nu is the Nusselt number with value of 1 for solid state and static liquid, and larger than 1 for liquid state under natural convection.

Considering the PCTD, the natural convection is prone to be induced by the top cooler and bottom heater, and the Nu of the natural convection is given as follows:[S4]

$$\begin{cases} Nu = 0.212(Gr_\delta Pr)^{1/4}, & 1.0 \times 10^4 \leq Gr_\delta \leq 4.6 \times 10^5 \\ Nu = 0.061(Gr_\delta Pr)^{1/3}, & Gr_\delta > 4.6 \times 10^5 \end{cases} \tag{S7}$$

Where, $Gr_\delta$ is the Grashof number and Pr is the Prandtl number of aqueous $CaCl_2$ solution in liquid phase, which are further determined by the following equations:

$$Gr_\delta = \frac{\rho^2 g \alpha_V \Delta T_{B2} \delta_B^3}{\mu^2} \tag{S8}$$

$$Pr = \frac{\mu c_p}{k_{Bl}} \tag{S9}$$

Where, $\rho$, $\alpha_V$, $c_p$ and $\mu$ are the density, thermal expansivity, specific heat capacity and viscosity of aqueous $CaCl_2$ solution, respectively; $\Delta T_{B2}$ is the temperature difference between two ends of aqueous $CaCl_2$ solution.

By following the mathematical models as given above, it is complicated to excatly estimate $k_{Bl\_eff}$ becuase many properties are required, such as thermal conductivity, density, thermal expansivity, specific heat capacity and viscosity.Therefore, a simple method was used to evaluate the $k_{Bl\_eff}$ by reorganizing Equation S6-S9 such that all the thermo-physical properties are combined into one parameter C, yielding a simplified equation as follows:

$$k_{Bl\_eff} = C \delta_B^{3/4} \Delta T_{Bl}^{1/4} \tag{S10}$$

Where, $C$ is expressed as follows:

$$C = 0.212 \left( \frac{\rho^2 g \alpha_V c_p k_{Bl}^3}{\mu} \right)^{1/4} \tag{S11}$$

When the mean temperature of aqueous $CaCl_2$ solution is fixed, the variation of $k_{Bl\_eff}$ is only related to the temperature difference $\Delta T_{B2}$ and the height of aqueous $CaCl_2$ solution $\delta_B$. Meanwhile, the parameter $C$ can be presented by a reference effective thermal conductivity $k_{Bl\_eff}^*$ of aqueous $CaCl_2$ solution at a reference temperature difference of $\Delta T_{B1}^*$ and reference liquid height of $\delta_B^*$:

$$C = (\frac{1}{\delta_B^*})^{3/4} (\frac{1}{\Delta T_{B1}^*})^{1/4} k_{Bl\_eff}^* \tag{S12}$$

Where, $k_{Bl\_eff}$ at an arbitrary $\Delta T_{B2}$ and $\delta_B$ is formulated as follows:

$$k_{Bl\_eff} = (\frac{\delta_B}{\delta_B^*})^{3/4} (\frac{\Delta T_{B1}}{\Delta T_{B1}^*})^{1/4} k_{Bl\_eff}^* \tag{S13}$$

In the present study, by defining the reference temperature difference $\Delta T_{B2}^*=10$ °C and the reference liquid height of $\delta_B^*=40$ mm, the effective thermal conductivities $k_{Bl\_eff}^*$ of aqueous $CaCl_2$ solution at different temperatures were measured by the steady-state method introduced in Section 2.1 of SI. Accordingly, $k_{Bl\_eff}^*$ of hot aqueous $CaCl_2$ solution came out to be 4.76 W m$^{-1}$ K$^{-1}$ at a mean temperature of 35 °C, and that of supercooled aqueous $CaCl_2$ solution was measured to be 3.15 W m$^{-1}$ K$^{-1}$ at a mean temperature of 20 °C. Therefore, the influence of natural convection of aqueous $CaCl_2$ solution can be depicted by combining the value of $k_{Bl\_eff}^*$ and Equation S13.

3. Mathematical modeling of the PCTD

A theoretical model is established that has a verstality to systematically address heat conduction and convection phenomena. Unlike the previous models as reported in references [20] [21] [37] (cited in the main text), the present model can solve the problem statements related to solid-solid and solid-liquid phase change phenomena, as well as help configure eighteen thermal modules for heating and cooling processes. Also, the limitation of constant thermal conductivities [21] [37] and constant effective thermal conductivities [20] is resolved in the present study, and theoretical solutions can be achieved with variable effective thermal conductivity in terms of liquid height. Since the PCTD undergoes an extended convection due to supercooling effect of aqueous $CaCl_2$ solution, a novel parameter $T_{fB}$ is included that ensures multiple benefits, including the unprecedentedly high thermal rectification ratio, a widened temperature range, a sustainable performance and a large tunability. Further detail of the procedure can be understood as follows:

- In the heating process, the phase change temperatures of PCM A (paraffin) and PCM B ($CaCl_2 \cdot 6H_2O$) were $T_{mA}$ and $T_{mB}$, respectively.
- In the cooling process, two approaches have been executed: before supercooling release, the phase change temperatures of PCM A and B were $T_{mA}$ and $T_{fB}$, respectively; after supercooling release, phase change temperatures of PCM A and B were recovered to $T_{mA}$ and $T_{mB}$, respectively, due to supercooling eliminated in liquid phase..

3.1 Heating process

In heating process, four thermal modules in forward direction and four in reverse direction have been modeled by relating the values of $T_i$, $T_{mA}$ and $T_{mB}$, as shown in Table S1. $\delta$ represents the liquid height of the corresponding PCMs throughout the modeling.

Table S1. Schematic illustration of different theoretical PCTD modules designed via possible operating conditions in heating process. The dark blue and light blue colors denote PCM A in solid state and liquid state, whose thermal conductivities are $k_{As}$ and $k_{Al}$; while dark red and light red colors represent PCM B in solid state and liquid state, whose effective thermal conductivities are $k_{Bs}$ and $k_{Bl\_eff}$.

| Operating Conditions | $T_i > T_{mA}$ and $T_i \geq T_{mB}$ | $T_{mB} \leq T_i \leq T_{mA}$ | $T_{mA} < T_i < T_{mB}$ | $T_i \leq T_{mA}$ and $T_i < T_{mB}$ |
|---|---|---|---|---|
| Forward Direction | Forward-H1 | Forward-H2 | Forward-H3 | Forward-H4 |
| Operating Conditions | $T_i \geq T_{mA}$ and $T_i > T_{mB}$ | $T_{mB} < T_i < T_{mA}$ | $T_{mA} \leq T_i \leq T_{mB}$ | $T_i < T_{mA}$ and $T_i \leq T_{mB}$ |
| Reverse Direction | Reverse-H1 | Reverse-H2 | Reverse-H3 | Reverse-H4 |

Each thermal module can be depicted at a certain operating condition with respective thermal resistances in series. According to the Fourier's law of heat conduction, the equations to depict the heat transfer of the PCTD at steady state are mathematically formulated as follows:

$$\begin{cases} \dfrac{\partial}{\partial x}(k_{As} \dfrac{\partial T}{\partial x}) = 0 \\ \dfrac{\partial}{\partial x}(k_{Al} \dfrac{\partial T}{\partial x}) = 0 \\ \dfrac{\partial}{\partial x}(k_{Bs} \dfrac{\partial T}{\partial x}) = 0 \\ \dfrac{\partial}{\partial x}(k_{Bl\_eff} \dfrac{\partial T}{\partial x}) = 0 \end{cases} \tag{S14}$$

Since the heat conduction of the PCTD is unidirectional, the heat fluxes through different junctions of the PCTD are equal, which is shown as follows:

$$q = -k_{As} \dfrac{\partial T}{\partial x} = -k_{Al} \dfrac{\partial T}{\partial x} = -k_{Bs} \dfrac{\partial T}{\partial x} = -k_{Bl\_eff} \dfrac{\partial T}{\partial x} \tag{S15}$$

Owing to different physical states of two junctions, the formulae of the temperature distributions and heat fluxes of the PCTD in different operating conditions are also different, which are listed as follows:

**Forward-H1:**

$$\begin{cases} T_{As}(x) = -\dfrac{T_{mA} - T_c}{L_A - \delta_A}(x - \delta_A - L_B) + T_{mA} \\ T_{Al}(x) = -\dfrac{T_i - T_{mA}}{\delta_A}(x - L_B) + T_i \\ T_{Bl}(x) = -\dfrac{T_h - T_i}{L_B}x + T_h \end{cases} \tag{S16}$$

$$q = \dfrac{T_{mA} - T_c}{L_A - \delta_A} k_{As} = \dfrac{T_i - T_{mA}}{\delta_A} k_{Al} = \dfrac{T_h - T_i}{L_B} k_{Bl\_eff} \tag{S17}$$

Where, $q$, $T_i$, $\delta_A$ and $\delta_B$ can be derived from Equations S16 and S17, and the formulae are given as below:

$$\begin{cases} T_i = -\dfrac{k_{Al}(T_h - T_{mA}) + k_{As}(T_{mA} - T_c)}{k_{Al}L_B + k_{Bl\_eff}L_A} L_B + T_h \\ \delta_A = \dfrac{k_{Al}}{k_{Bl\_eff}} \dfrac{k_{Bl\_eff}L_A(T_h - T_{mA}) - k_{As}L_B(T_{mA} - T_c)}{k_{Al}(T_h - T_{mA}) + k_{As}(T_{mA} - T_c)} \\ \delta_B = L_B \\ q = k_{Bl\_eff} \dfrac{k_{Al}(T_h - T_{mA}) + k_{As}(T_{mA} - T_c)}{k_{Al}L_B + k_{Bl\_eff}L_A} \end{cases} \tag{S18}$$

**Forward-H2:**

$$\begin{cases} T_{As}(x) = -\dfrac{T_i - T_c}{L_A}(x - L_B) + T_i \\ T_{Bl}(x) = -\dfrac{T_h - T_i}{L_B}x + T_h \end{cases} \tag{S19}$$

$$q = \dfrac{T_i - T_c}{L_A} k_{As} = \dfrac{T_h - T_i}{L_B} k_{Bl\_eff} \tag{S20}$$

Where, $q$, $T_i$, $\delta_A$ and $\delta_B$ can be derived from Equations S19 and S20, and the formulae are given as below:

$$\begin{cases} T_i = \dfrac{k_{Bl\_eff}L_A T_h + k_{As}L_B T_c}{k_{As}L_B + k_{Bl\_eff}L_A} \\ \delta_A = 0 \\ \delta_B = L_B \\ q = \dfrac{k_{As}k_{Bl\_eff}(T_h - T_c)}{k_{As}L_B + k_{Bl\_eff}L_A} \end{cases} \tag{S21}$$

**Forward-H3:**

$$\begin{cases} T_{As}(x) = -\dfrac{T_{mA} - T_c}{L_A - \delta_A}(x - L_B - \delta_A) + T_{mA} \\ T_{Al}(x) = -\dfrac{T_i - T_{mA}}{\delta_A}(x - L_B) + T_i \\ T_{Bs}(x) = -\dfrac{T_{mB} - T_i}{L_B - \delta_B}(x - \delta_B) + T_{mB} \\ T_{Bl}(x) = -\dfrac{T_h - T_{mB}}{\delta_B}x + T_h \end{cases} \tag{S22}$$

$$q = \dfrac{T_{mA} - T_c}{L_A - \delta_A}k_{As} = \dfrac{T_i - T_{mA}}{\delta_A}k_{Al} = \dfrac{T_{mB} - T_i}{L_B - \delta_B}k_{Bs} = \dfrac{T_h - T_{mB}}{\delta_B}k_{Bl\_eff} \tag{S23}$$

Where, $q$, $T_i$, $\delta_A$ and $\delta_B$ can be derived from Equations S22 and S23, and the formulae are given as below:

$$\begin{cases} T_i = \dfrac{k_{Bl\_eff}L_A(T_h - T_{mB}) + k_{Bs}L_A(T_{mB} - T_{mA}) - k_{As}L_B(T_{mA} - T_c)}{k_{Al}L_B + k_{Bs}L_A} + T_{mA} \\ \delta_A = \dfrac{k_{Bl\_eff}L_A(T_h - T_{mB}) + k_{Bs}L_A(T_{mB} - T_{mA}) - k_{As}L_B(T_{mA} - T_c)}{k_{Al}k_{Bl\_eff}(T_h - T_{mB}) + k_{Al}k_{Bs}(T_{mB} - T_{mA}) + k_{As}k_{Bs}(T_{mA} - T_c)}k_{Al} \\ \delta_B = \dfrac{(k_{Al}L_B + k_{Bs}L_A)(T_h - T_{mB})}{k_{Al}k_{Bl\_eff}(T_h - T_{mB}) + k_{Al}k_{Bs}(T_{mB} - T_{mA}) + k_{As}k_{Bs}(T_{mA} - T_c)}k_{Bl\_eff} \\ q = \dfrac{k_{Al}k_{Bl\_eff}(T_h - T_{mB}) + k_{Al}k_{Bs}(T_{mB} - T_{mA}) + k_{As}k_{Bs}(T_{mA} - T_c)}{k_{Al}L_B + k_{Bs}L_A} \end{cases} \tag{S24}$$

**Forward-H4:**

$$\begin{cases} T_{As}(x) = -\dfrac{T_i - T_c}{L_A}(x - L_B) + T_i \\ T_{Bs}(x) = -\dfrac{T_{mB} - T_i}{L_B - \delta_B}(x - \delta_B) + T_{mB} \\ T_{Bl}(x) = -\dfrac{T_h - T_{mB}}{\delta_B}x + T_h \end{cases} \tag{S25}$$

$$q = \dfrac{T_i - T_c}{L_A}k_{As} = \dfrac{T_{mB} - T_i}{L_B - \delta_B}k_{Bs} = \dfrac{T_h - T_{mB}}{\delta_B}k_{Bl\_eff} \tag{S26}$$

Where, $q$, $T_i$, $\delta_A$ and $\delta_B$ can be derived from Equations S25 and S26, and the formulae are given as below:

$$\begin{cases} T_i = \dfrac{k_{Bl\_eff}(T_h - T_{mB}) + k_{Bs}(T_{mB} - T_c)}{k_{As}L_B + k_{Bs}L_A} L_A + T_c \\ \delta_A = 0 \\ \delta_B = \dfrac{k_{Bl\_eff}}{k_{As}} \dfrac{(k_{As}L_B + k_{Bs}L_A)(T_h - T_{mB})}{k_{Bl\_eff}(T_h - T_{mB}) + k_{Bs}(T_{mB} - T_c)} \\ q = k_{As} \dfrac{k_{Bl\_eff}(T_h - T_{mB}) + k_{Bs}(T_{mB} - T_c)}{k_{As}L_B + k_{Bs}L_A} \end{cases} \quad (S27)$$

**Reverse-H1:**

$$\begin{cases} T_{Al}(x) = -\dfrac{T_h - T_i}{L_A} x + T_h \\ T_{Bs}(x) = -\dfrac{T_{mB} - T_c}{L_B - \delta_B}(x - L_A - \delta_B) + T_{mB} \\ T_{Bl}(x) = -\dfrac{T_i - T_{mB}}{\delta_B}(x - L_A) + T_i \end{cases} \quad (S28)$$

$$q = \dfrac{T_h - T_i}{L_A} k_{Al} = \dfrac{T_{mB} - T_c}{L_B - \delta_B} k_{Bs} = \dfrac{T_i - T_{mB}}{\delta_B} k_{Bl\_eff} \quad (S29)$$

Where, $q$, $T_i$, $\delta_A$ and $\delta_B$ can be derived from Equations S28 and S29, and the formulae are given as below:

$$\begin{cases} T_i = -\dfrac{k_{Bl\_eff}(T_h - T_{mB}) + k_{Bs}(T_{mB} - T_c)}{k_{Al}L_B + k_{Bl\_eff}L_A} L_A + T_h \\ \delta_A = L_A \\ \delta_B = \dfrac{k_{Bl\_eff}}{k_{Al}} \dfrac{k_{Al}L_B(T_h - T_{mB}) - k_{Bs}L_A(T_{mB} - T_c)}{k_{Bl\_eff}(T_h - T_{mB}) + k_{Bs}(T_{mB} - T_c)} \\ q = k_{Al} \dfrac{k_{Bl\_eff}(T_h - T_{mB}) + k_{Bs}(T_{mB} - T_c)}{k_{Al}L_B + k_{Bl\_eff}L_A} \end{cases} \quad (S30)$$

**Reverse-H2:**

$$\begin{cases} T_{As}(x) = -\dfrac{T_{mA} - T_i}{L_A - \delta_A}(x - \delta_A) + T_{mA} \\ T_{Al}(x) = -\dfrac{T_h - T_{mA}}{\delta_A} x + T_h \\ T_{Bs}(x) = -\dfrac{T_{mB} - T_c}{L_B - \delta_B}(x - L_A - \delta_B) + T_{mB} \\ T_{Bl}(x) = -\dfrac{T_i - T_{mB}}{\delta_B}(x - L_A) + T_i \end{cases} \quad (S31)$$

$$q = \frac{T_{mA} - T_i}{L_A - \delta_A} k_{As} = \frac{T_h - T_{mA}}{\delta_A} k_{Al} = \frac{T_{mB} - T_c}{L_B - \delta_B} k_{Bs} = \frac{T_i - T_{mB}}{\delta_B} k_{Bl\_eff} \quad (S32)$$

Where, $q$, $T_i$, $\delta_A$ and $\delta_B$ can be derived from Equations S31 and S32, and the formulae are given as below:

$$\begin{cases} T_i = \dfrac{k_{Al}L_B(T_h - T_{mA}) + k_{As}L_B(T_{mA} - T_{mB}) - k_{Bs}L_A(T_{mB} - T_c)}{k_{As}L_B + k_{Bl\_eff}L_A} + T_{mB} \\[2mm] \delta_A = \dfrac{(k_{As}L_B + k_{Bl\_eff}L_A)(T_h - T_{mA})}{k_{Al}k_{Bl\_eff}(T_h - T_{mA}) + k_{As}k_{Bl\_eff}(T_{mA} - T_{mB}) + k_{As}k_{Bs}(T_{mB} - T_c)} k_{Al} \\[2mm] \delta_B = \dfrac{k_{Al}L_B(T_h - T_{mA}) + k_{As}L_B(T_{mA} - T_{mB}) - k_{Bs}L_A(T_{mB} - T_c)}{k_{Al}k_{Bl\_eff}(T_h - T_{mA}) + k_{As}k_{Bl\_eff}(T_{mA} - T_{mB}) + k_{As}k_{Bs}(T_{mB} - T_c)} k_{Bl\_eff} \\[2mm] q = \dfrac{k_{Al}k_{Bl\_eff}(T_h - T_{mA}) + k_{As}k_{Bl\_eff}(T_{mA} - T_{mB}) + k_{As}k_{Bs}(T_{mB} - T_c)}{k_{As}L_B + k_{Bl\_eff}L_A} \end{cases} \quad (S33)$$

**Reverse-H3:**

$$\begin{cases} T_{Al}(x) = -\dfrac{T_h - T_i}{L_A} x + T_h \\[2mm] T_{Bs}(x) = -\dfrac{T_i - T_c}{L_B}(x - L_B) + T_i \end{cases} \quad (S34)$$

$$q = \frac{T_h - T_i}{L_A} k_{Al} = \frac{T_i - T_c}{L_B} k_{Bs} \quad (S35)$$

Where, $q$, $T_i$, $\delta_A$ and $\delta_B$ can be derived from Equations S34 and S35, and the formulae are given as below:

$$\begin{cases} T_i = \dfrac{k_{Al}L_B T_h + k_{Bs}L_A T_c}{k_{Al}L_B + k_{Bs}L_A} \\[2mm] \delta_A = L_A \\ \delta_B = 0 \\[2mm] q = \dfrac{k_{Al}k_{Bs}(T_h - T_c)}{k_{Al}L_B + k_{Bs}L_A} \end{cases} \quad (S36)$$

**Reverse-H4:**

$$\begin{cases} T_{As}(x) = -\dfrac{T_{mA} - T_i}{L_A - \delta_A}(x - \delta_A) + T_{mA} \\[2mm] T_{Al}(x) = -\dfrac{T_h - T_{mA}}{\delta_A} x + T_h \\[2mm] T_{Bs}(x) = -\dfrac{T_i - T_c}{L_B}(x - L_A) + T_i \end{cases} \quad (S37)$$

$$q = \frac{T_{mA} - T_i}{L_A - \delta_A} k_{As} = \frac{T_h - T_{mA}}{\delta_A} k_{Al} = \frac{T_i - T_c}{L_B} k_{Bs} \tag{S38}$$

Where, $q$, $T_i$, $\delta_A$ and $\delta_B$ can be derived from Equations S37 and S38, and the formulae are given as below:

$$\begin{cases} T_i = \dfrac{k_{Al} L_B (T_h - T_{mA}) - k_{Bs} L_A (T_{mA} - T_c)}{k_{As} L_B + k_{Bs} L_A} + T_{mA} \\ \delta_A = \dfrac{k_{Al}}{k_{Bs}} \dfrac{(k_{As} L_B + k_{Bs} L_A)(T_h - T_{mA})}{k_{Al}(T_h - T_{mA}) + k_{As}(T_{mA} - T_c)} \\ \delta_B = 0 \\ q = k_{Bs} \dfrac{k_{Al}(T_h - T_{mA}) + k_{As}(T_{mA} - T_c)}{k_{As} L_B + k_{Bs} L_A} \end{cases} \tag{S39}$$

Using the formulated heat fluxes of the PCTD in different operating conditions, thermal rectification ratio in each thermal module of the PCTD can be calculated by dividing forward heat flux to reverse heat flux.

3.2 Cooling Process

In cooling process, the theoretical model of the PCTD introduced in Section 3.1 of SI is modified. The supercooling of PCM B releases spontaneously at a temperature, denoted as $T_{fB}$, which is a key parameter playing an important role in the modified theoretical model. In forward direction, PCM B is near the bottom heater, and the coldest part of PCM B is at the interface of PCM A and PCM B, where the temperature is $T_i$. There are four operating conditions categorized by the relationships among the values of $T_i$, $T_{mA}$ and $T_{fB}$. When $T_i > T_{fB}$, PCM B is in liquid and supercooled liquid phase; when $T_i \leq T_{fB}$, the supercooling of PCM B releases spontaneously and part of PCM B exists in solid phase, where the location of the solid-liquid interface of PCM B is determined by $T_{mB}$. In reverse direction, PCM B is adjacent to the top cooler with the coldest part at the top, where the temperature is $T_c$. The states of PCM A and PCM B are determined by the relationship among the values of $T_i$, $T_c$, $T_{mA}$, $T_{fB}$ and $T_{mB}$. Consequently, there are totally six different operating conditions in reverse direction.

Table S2. Schematic illustration of different theoretical PCTD modules designed via possible operating conditions in cooling process. The color of dark blue and light blue exhibit PCM A in solid state and liquid state, whose thermal conductivities are $k_{As}$ and $k_{Al}$, respectively; while dark red and light red show PCM B in solid state and liquid state, whose effective thermal conductivities are $k_{Bs}$ and $k_{Bl\_eff}$, respectively.

| Operating Conditions | $T_i>T_{mA}$ and $T_i>T_{fB}$ | $T_{fB}<T_i\leq T_{mA}$ | $T_{mA}<T_i\leq T_{fB}$ | $T_i\leq T_{mA}$ and $T_i\leq T_{fB}$ | | |
|---|---|---|---|---|---|---|
| Forward Direction | Forward-C1 | Forward-C2 | Forward-C3 | Forward-C4 | | |
| Operating Conditions | $T_i\geq T_{mA}$ and $T_c>T_{fB}$ | $T_i<T_{mA}$ and $T_c>T_{fB}$ | $T_i\geq T_{mA}$, $T_i>T_{mB}$ and $T_c\leq T_{fB}$ | $T_i<T_{mA}$, $T_i>T_{mB}$ and $T_c\leq T_{fB}$ | $T_i\geq T_{mA}$, $T_i\leq T_{mB}$ and $T_c\leq T_{fB}$ | $T_i<T_{mA}$, $T_i\leq T_{mB}$ and $T_c\leq T_{fB}$ |
| Reverse Direction | Reverse-C1 | Reverse-C2 | Reverse-C3 | Reverse-C4 | Reverse-C5 | Reverse-C6 |

In cooling process, four thermal modules in forward direction and six in reverse direction have been modeled, as shown in Table S2. Each thermal module can be depicted at a certain operating condition with respective thermal resistances in series. Furthermore, a majority of thermal modules of the PCTD in cooling process are identical to those in heating process because of the same physical states. For instance, four thermal modules of forward direction (Forward-C1 to C4) and four of reverse direction (Reverse-C3 to C6) in cooling process are the same as those of forward direction (Forward-H1 to H4) and reverse direction (Reverse-H1 to H4) in heating process, respectively. Therefore, the equations for Forward-C1 to Forward-C4 and Reverse-C3 to Reverse-C6 have been adopted from their similar counterparts. Subsequently, the calculation procedures of remaining two thermal modules, Reverse-C1 and Reverse-C2, are introduced as follows:

**Reverse-C1:**

$$\begin{cases} T_{Al}(x) = -\dfrac{T_h - T_i}{L_A} x + T_h \\ T_{Bl}(x) = -\dfrac{T_i - T_c}{L_B}(x - L_B) + T_i \end{cases} \tag{S40}$$

$$q = \dfrac{T_h - T_i}{L_A} k_{Al} = \dfrac{T_i - T_c}{L_B} k_{Bl\_eff} \tag{S41}$$

Where, $q$, $T_i$, $\delta_A$ and $\delta_B$ can be inferred from Equations S40 and S41, and the formulae are given as below:

$$\begin{cases} T_i = \dfrac{k_{Al} L_B T_h + k_{Bl\_eff} L_A T_c}{k_{Al} L_B + k_{Bl\_eff} L_A} \\ \delta_A = L_A \\ \delta_B = L_B \\ q = \dfrac{k_{Al} k_{Bl\_eff}(T_h - T_c)}{k_{Al} L_B + k_{Bl\_eff} L_A} \end{cases} \tag{S42}$$

**Reverse-C2:**

$$\begin{cases} T_{As}(x) = -\dfrac{T_{mA} - T_i}{L_A - \delta_A}(x - \delta_A) + T_{mA} \\ T_{Al}(x) = -\dfrac{T_h - T_{mA}}{\delta_A} x + T_h \\ T_{Bl}(x) = -\dfrac{T_i - T_c}{L_B}(x - L_A) + T_i \end{cases} \tag{S43}$$

$$q = \frac{T_{mA} - T_i}{L_A - \delta_A} k_{As} = \frac{T_h - T_{mA}}{\delta_A} k_{Al} = \frac{T_i - T_c}{L_B} k_{Bl\_eff} \tag{S44}$$

Where, $q$, $T_i$, $\delta_A$ and $\delta_B$ can be inferred from Equations S43 and S44, and the formulae are given as below:

$$\begin{cases} T_i = \dfrac{k_{Al} L_B (T_h - T_{mA}) - k_{Bl\_eff} L_A (T_{mA} - T_c)}{k_{As} L_B + k_{Bl\_eff} L_A} + T_{mA} \\ \delta_A = \dfrac{k_{Al}}{k_{Bl\_eff}} \dfrac{(T_h - T_{mA})(k_{As} L_B + k_{Bl\_eff} L_A)}{k_{Al}(T_h - T_{mA}) + k_{As}(T_{mA} - T_c)} \\ \delta_B = L_B \\ q = k_{Bl\_eff} \dfrac{k_{Al}(T_h - T_{mA}) + k_{As}(T_{mA} - T_c)}{k_{As} L_B + k_{Bl\_eff} L_A} \end{cases} \tag{S45}$$

4. Validation of mathematical model of the PCTD under influence of thermal contact resistance

To theoretically describe the experiments in a more reasonable manner, the influence of thermal contact resistance on the heat flux of the PCTD should be taken into consideration. The temperature distribution in the PCTD and the states of two PCM terminals are influenced by thermal contact resistance, and the heat flux of the PCTD is accordingly subjected to a slight change. Therefore, the theoretical modules in Section 3 of SI should be modified to include the thermal contact resistance in the 1-D heat conduction model, which is called comprehensive method. We take the most complicated theoretical module (Forward-C3 in Section 3.2 of SI) as an example to explain the procedure:

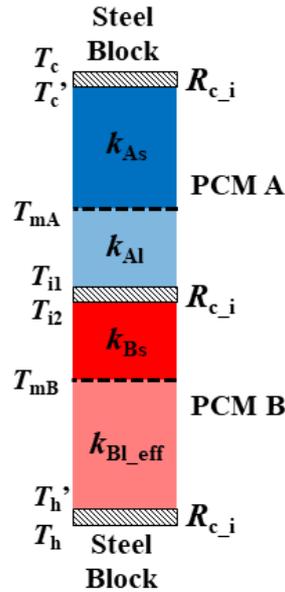

Figure S3. Schematic illustration of one example of theoretical PCTD module (Forward-C3 in Section 3.2 of SI) that takes the influence of the thermal contact resistance into consideration.

The dark blue and light blue colors illustrate PCM A in solid and liquid states, whose thermal conductivities are $k_{As}$ and $k_{Al}$, respectively; while dark red and light red show PCM B in solid and liquid states, whose thermal conductivities are $k_{Bs}$ and $k_{Bl\_eff}$, respectively. In the figure, three segments of thermal contact resistances correspond to those between steel block and PCM B, PCM B and PCM A as well as PCM A and steel block from bottom to top.

There are three thermal contact resistances in the theoretical PCTD module, one is between the two PCM terminals, and the other two are between the PCTD and two steel blocks. As can be found form the results in Section 2.1 of SI, the values of different thermal contact resistances are almost the same because thermal grease is pasted between all the interfaces, which is denoted as $R_{c\_i}$. Consequently, the value of $R_{c\_i}=0.0015$ K m$^2$ W$^{-1}$ is one third of that of $R_{c\_td}$.

Owing to the presence of thermal contact resistance, there is a temperature difference across the interface. For the interface between PCM A and PCM B, the interface temperature is $T_{i1}$ on the side of PCM A, and $T_{i2}$ on the side of PCM B; for the interface between top steel block and PCM A, the interface temperature is $T_c$ on the side of top steel block, and $T_c'$ on the side of PCM A; for the interface between PCM B and bottom steel block, the interface temperature is $T_h$ on the side of bottom steel block, and $T_h'$ on the side of PCM A. There is a relationship between the interface temperatures and the thermal contact resistances:

$$q_c = \frac{T_c - T_c'}{R_{c\_i}} = \frac{T_{i1} - T_{i2}}{R_{c\_i}} = \frac{T_h - T_h'}{R_{c\_i}} = \frac{\Delta T_c}{R_{c\_i}} \tag{S46}$$

Where, $q_c$ is the heat flux of the PCTD, and $\Delta T_c$ is the temperature difference across the interface.

The above thermal module can be depicted by a 1-D heat conduction model with respective thermal resistances in series. According to the Fourier's law of heat conduction, the equations to depict the temperature distribution of the PCTD at steady state are mathematically formulated as follows:

$$\begin{cases} T_{As}(x) = -\dfrac{T_{mA} - T_c'}{L_A - \delta_A}(x - L_B - \delta_A) + T_{mA} \\[2mm] T_{Al}(x) = -\dfrac{T_{i1} - T_{mA}}{\delta_A}(x - L_B) + T_{i1} \\[2mm] T_{Bs}(x) = -\dfrac{T_{mB} - T_{i2}}{L_B - \delta_B}(x - \delta_B) + T_{mB} \\[2mm] T_{Bl}(x) = -\dfrac{T_h' - T_{mB}}{\delta_B}x + T_h' \end{cases} \tag{S47}$$

Since the heat conduction of the PCTD is unidirectional, the heat fluxes through different junctions of the PCTD are equal, which is shown as follows:

$$q_c = \frac{T_{mA} - T_c'}{L_A - \delta_A} k_{As} = \frac{T_{i1} - T_{mA}}{\delta_A} k_{Al} = \frac{T_{mB} - T_{i2}}{L_B - \delta_B} k_{Bs} = \frac{T_h' - T_{mB}}{\delta_B} k_{Bl\_eff} = \frac{\Delta T_c}{R_{c\_i}} \tag{S48}$$

Where, $q_c$, $T_{i1}$, $\Delta T_c$, $\delta_A$ and $\delta_B$ can be derived from Equations S46-S48, and the formulae are given as below:

$$\begin{cases} T_{i1} = \frac{k_{Al}k_{Bl\_eff}(T_h - T_{mA}) + k_{As}k_{Bl\_eff}(T_{mA} - T_{mB}) + k_{As}k_{Bs}(T_{mB} - T_c)}{k_{As}L_B + k_{Bl\_eff}L_A + k_{As}k_{Bs}R_{c\_i} + k_{As}k_{Bl\_eff}R_{c\_i} + k_{Al}k_{Bl\_eff}R_{c\_i}} \frac{L_B + k_{Bs}R_{c\_i}}{k_{Bl\_eff}} + T_{mB} - \frac{k_{Bs}}{k_{Bl\_eff}}(T_{mB} - T_c) \\ \Delta T_c = \frac{k_{Al}k_{Bl\_eff}(T_h - T_{mA}) + k_{As}k_{Bl\_eff}(T_{mA} - T_{mB}) + k_{As}k_{Bs}(T_{mB} - T_c)}{k_{As}L_B + k_{Bl\_eff}L_A + k_{As}k_{Bs}R_{c\_i} + k_{As}k_{Bl\_eff}R_{c\_i} + k_{Al}k_{Bl\_eff}R_{c\_i}} R_{c\_i} \\ \delta_A = \frac{k_{As}L_B + k_{Bl\_eff}L_A + k_{As}k_{Bs}R_{c\_i} + k_{As}k_{Bl\_eff}R_{c\_i} + k_{Al}k_{Bl\_eff}R_{c\_i}}{k_{Al}k_{Bl\_eff}(T_h - T_{mA}) + k_{As}k_{Bl\_eff}(T_{mA} - T_{mB}) + k_{As}k_{Bs}(T_{mB} - T_c)}(T_h - T_{mA})k_{Al} - k_{Al}R_{c\_i} \\ \delta_B = \frac{k_{As}L_B + k_{Bl\_eff}L_A + k_{As}k_{Bs}R_{c\_i} + k_{As}k_{Bl\_eff}R_{c\_i} + k_{Al}k_{Bl\_eff}R_{c\_i}}{k_{Al}k_{Bl\_eff}(T_h - T_{mA}) + k_{As}k_{Bl\_eff}(T_{mA} - T_{mB}) + k_{As}k_{Bs}(T_{mB} - T_c)}(T_c - T_{mB})k_{Bs} + L_B + k_{As}R_{c\_i} \\ q_c = \frac{k_{Al}k_{Bl\_eff}(T_h - T_{mA}) + k_{As}k_{Bl\_eff}(T_{mA} - T_{mB}) + k_{As}k_{Bs}(T_{mB} - T_c)}{k_{As}L_B + k_{Bl\_eff}L_A + k_{As}k_{Bs}R_{c\_i} + k_{As}k_{Bl\_eff}R_{c\_i} + k_{Al}k_{Bl\_eff}R_{c\_i}} \end{cases} \tag{S49}$$

And $T_{i2}$, $T_c'$ and $T_h'$ can be calculated by $T_{i2} = T_{i1} + \Delta T_c$, $T_c' = T_c + \Delta T_c$, $T_h' = T_h - \Delta T_c$, respectively.

The $q_c$ of other thermal modules can also be described by the comprehensive model and estimated in the same manner. However, including thermal contact resistance in the thermal module brings very large complexity in theoretical analysis. We therefore propose a lumped-parameter method to take the effect of thermal contact resistances into consideration where the total thermal contact resistance is treated as a modifying factor for heat flux obtained by 1-D heat conduction model in Section 3 of SI, as shown by Equation 5 in the main text. The difference between the heat fluxes calculated by two methods at the experimental conditions are shown in Figure S4 (a and b). It can be found from the figure that the maximum deviation is smaller than 3 %. Consequently, balancing the precision and complexity of theoretical model, the lumped-parameter method for heat flux correction is adopted in the present work.

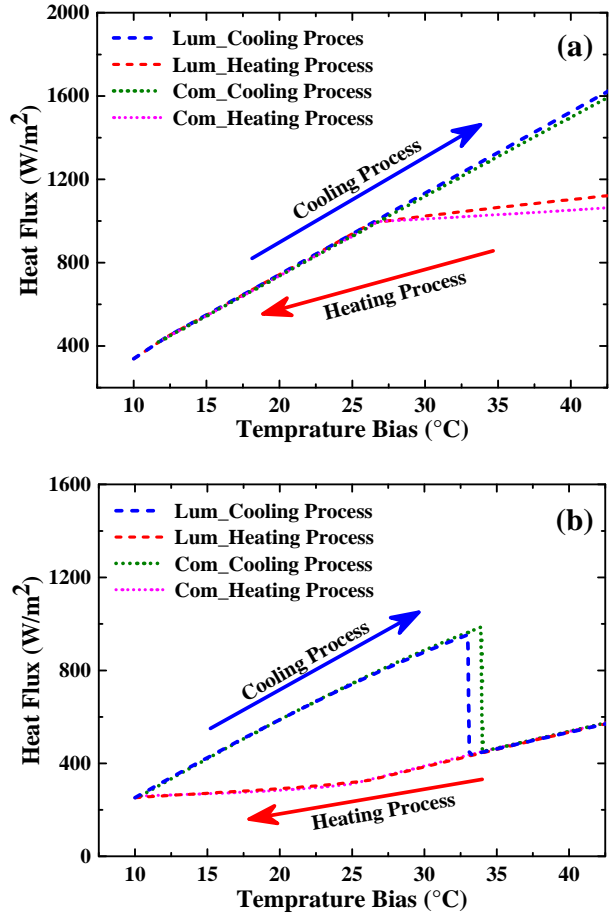

Figure S4. Numerical heat fluxes through the PCTD estimated by the lumped-parameter method (Lum) for heat flux correction and the comprehensive method (Com). (a) In the forward direction, the range of temperature bias is 10~40 °C, and the maximum deviation between heat fluxes of two methods is smaller than 3 %, which is at the temperature bias of 40 °C in the heating process. (b) In the reverse direction, the range of temperature bias is also 10~40 °C, and the temperature biases of spontaneous supercooling release of two methods have a slight difference (within 1 °C), which is attributed to the temperature deviation of $T_c$ and $T_c$' in the comprehensive method.

5. Fabrication of the PCTD prototype and experimental procedure

As depicted schematically in Figure S5a, two cylindrical containers made of polymethyl methacrylate with different heights were employed. Paraffin and $CaCl_2 \cdot 6H_2O$ were kept in the containers separately, which were named as paraffin terminal and $CaCl_2$ terminal of the PCTD.

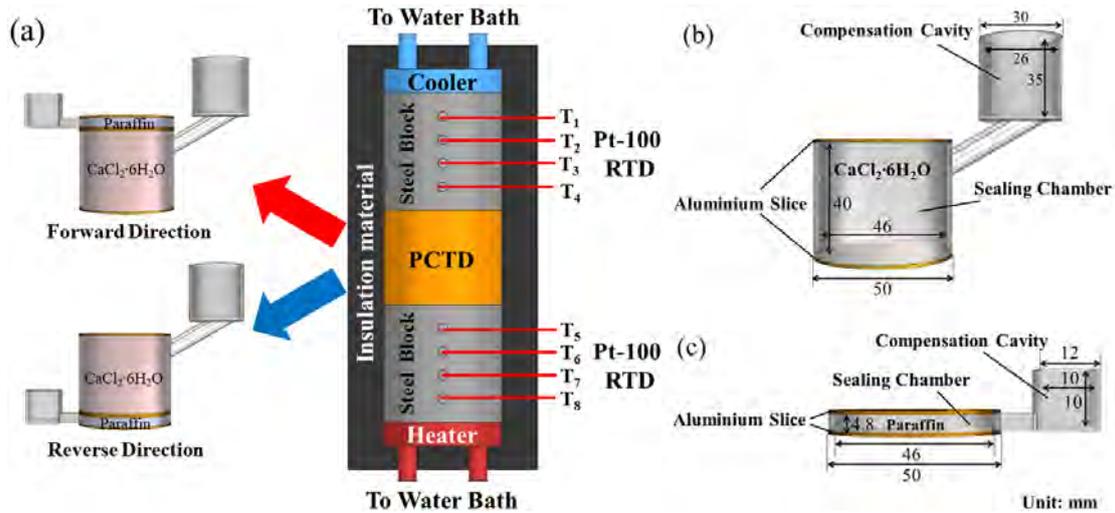

Figure S5. (a) Alternative terminals of the PCTD depicting the forward and reverse directions (left of a) and their placement within experimental set-up for heat flux determination (right of a). Dimensions of (b) $CaCl_2$ terminal, and (c) paraffin terminal along with compensation cavities. All units are in mm.

The dimensions of terminals were selected after getting systematic guidelines from theoretical analysis. The wall thickness of each terminal was 2.5 mm and outer diameter was 50 mm, while the heights of paraffin and $CaCl_2$ terminals came out to be 4.8 mm and 40 mm, respectively, as shown in Figure S5b and S5c. The top and bottom faces of terminals were closed by clamping 1 mm thick aluminum disk through thermal adhesive. To configure the PCTD working in the forward and reverse directions, paraffin and $CaCl_2$ terminals can be exchanged with each other, indicating the ease of operation as well as design scalability. For the forward direction operation, $CaCl_2$ terminal was positioned as the bottom terminal with paraffin as the top terminal, and for reverse direction operation, both terminals were switched. The alternative placement of both terminals is illustrated in left schematics of Figure S5a. To measure the heat flux across the fabricated PCTD, it was placed in the same experimental prototype where thermal conductivity of $CaCl_2·6H_2O$ was determined.

The major advantage of the presented configuration is the inclusion of lateral cavities, known as compensation cavities, which helps mitigate the risks associated with thermal contraction and thermal expansion of PCMs. Except than vulnerability of leakage in liquid state, inherent volumetric expansion and contraction of paraffin and $CaCl_2·6H_2O$ pose serious threats especially during solidification, which often generates a gap between PCM body and top aluminum cover of the container, hence destroying the junctional thermal contact and adversely affecting the thermal performance of the PCTD. To eliminate the influence of continuous volumetric variations, the compensation cavities were constructed and linked onto the top corner of cylindrical containers through an arm-like hollow channel as shown in Figure S5. As long as the melting prevails, the liquid PCM flows upward through the channel and tends to stay in the cavity until the solidification approaches during which liquid shrinks and flows back to the chamber under gravitational influence, thus keeping the chamber full of PCM. Therefore, during cyclic experimental operations continuing for several hours, volumetric expansion was successfully avoided. With addressing those risks, the expectations of having a contact loss between thermal media and boundaries of enclosure have been entirely depleted.

Based on the initial physical states of thermal media, the experiment was performed in two procedures, i.e., the first refers to the cooling process, and the second refers to the heating process. The entire procedures in the forward direction are explained in Figure S6, and the procedures in the reverse direction are identical to that in the forward direction. In the cooling process, aqueous $CaCl_2$ solution tended to be in liquid state so that supercooling was investigated both experimentally and theoretically. In heating process, solid $CaCl_2 \cdot 6H_2O$ melted gradually due to the high temperature of bottom heater and the influence of supercooling did not take place.

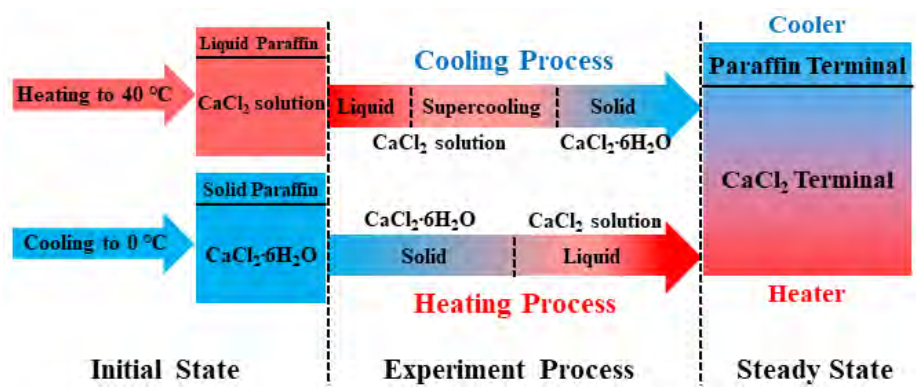

Figure S6. Schematic of state variation of $CaCl_2$ terminal in experimental procedure in the forward direction of the cooling and heating processes.

6. Method of manual supercooling release

Manual supercooling release (MSR) refers to elimination of supercooling by externally seeding the solid crystals into aqueous $CaCl_2$ solution.[S5] Influenced by phase change hysteresis, $CaCl_2$ solution retains liquid state in the temperature range of 7~30 °C, which is called supercooling zone. As schematically shown in Figure S7, by seeding $CaCl_2 \cdot 6H_2O$ crystal into aqueous $CaCl_2$ solution from compensation cavity, the supercooling of aqueous $CaCl_2$ solution was released, and the aqueous $CaCl_2$ solution in supercooling zone was converted to solid state instantaneously. The variation in the physical state influenced thermal rectification ratio of the PCTD, and a tunable PCTD can be achieved by manual supercooling release.

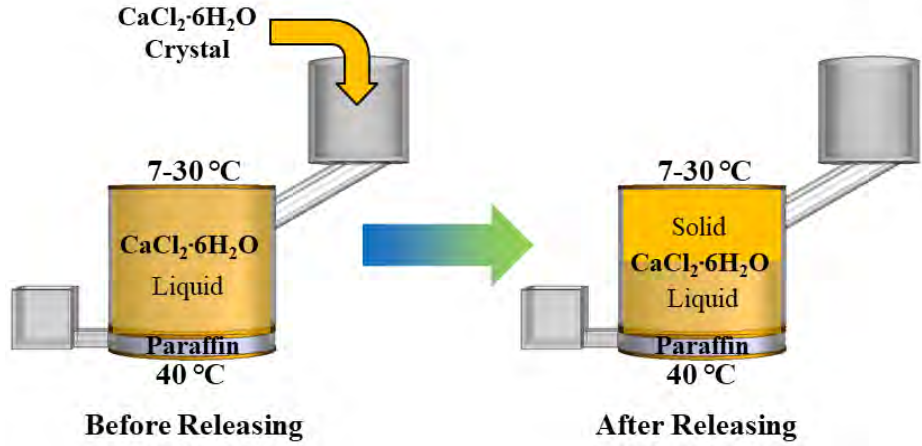

Figure S7. Manual supercooling release by seeding CaCl$_2$·6H$_2$O solid crystal into aqueous CaCl$_2$ solution.

7. Parametric uncertainty analysis

The uncertainty of the heat flux $q$ and thermal rectification ratio $R$ of the PCTD needs to be determined. In the present experiment, the heat flux of the PCTD was evaluated by the temperature gradient and thermal conductivity of steel blocks, which is shown in Equation S2. Furthermore, although the experimental set-up was encased by thermal insulation materials, the heat loss in experiments could not be completely eliminated, which can be evaluated by the difference between the heat fluxes of two steel blocks. Therefore, the major uncertainty in measurement was caused by the uncertainties of the temperature differences $\Delta T$ measured by the RTDs, the distances $\Delta x$ between the RTDs, the thermal conductivity $k_{st}$ of the steel block and the difference between the heat fluxes of two steel blocks $q_{st1}$ and $q_{st2}$, which are shown in Table S3. The measured uncertainty of heat flux can be estimated to be:

$$\delta q = \sqrt{(\delta k_{st} \frac{\Delta T}{\Delta x})^2 + (k_{st} \frac{\delta \Delta T}{\Delta x})^2 + (k_{st} \frac{\Delta T}{(\Delta x)^2} \delta \Delta x)^2 + \left|\frac{q_{st1} - q_{st2}}{q_{st1}}\right|^2} \quad (S50)$$

where $\delta$ presents the uncertainty of each parameter. The uncertainty of thermal rectification ratio can be evaluated by the uncertainties of the forward heat flux $q_{for}$ and the reverse heat flux $q_{rev}$:

$$\delta R = \sqrt{(\frac{\delta q_{for}}{q_{for}})^2 + (\frac{\delta q_{rev}}{q_{rev}})^2} \quad (S52)$$

The values of the uncertainty in heat flux and thermal rectification ratio of the PCTD under different conditions were evaluated, and the maximum uncertainty of heat flux is 30.85 % and the maximum uncertainty of thermal rectification ratio is 36.87 %.

Table S3. Results of parametric uncertainty analysis.

| Parameter | Uncertainty |
| --- | --- |
| Measured parameters | |
| Distance between RTDs, $x$ (m) | 0.001 |
| Temperature, $T$ (°C) | 0.1 |
| Thermal conductivity of steel, $k_{st}$ (W m$^{-1}$ K$^{-1}$) | 0.1 |
| Derived parameters | |
| Heat flux, $q$ (%) | 30.85 |
| Thermal rectification ratio, $R$ (%) | 36.87 |